\documentclass[twocolumn,amssymb,prb]{revtex4}
\usepackage{epsfig}
\usepackage{graphicx}

\begin{document}
\title{Properties of the Majorana-state tunneling Josephson junction mediated by an interacting quantum dot}

\author{Piotr Stefa\'nski}
\email{piotrs@ifmpan.poznan.pl}
 \affiliation{Institute of Molecular Physics of the Polish Academy of Sciences\\
  ul. Smoluchowskiego 17, 60-179 Pozna\'n, Poland}
\author{}
\affiliation{}
\author{}
\affiliation{}
\date{\today}

\begin{abstract}
We consider a model of a Josephson junction of two topological superconducting wires mediated by an interacting quantum dot. An additional normal electrode coupled to the dot from the top allows to probe its density of states. The Majorana states adjacent to the dot hybridize across the junction and from a bound state in the dot. The dot is subjected to the effective magnetic field arising from the superposition of the fields driving each wire into topological states, which, dependent on the angle between the fields, introduces variable Zeeman splitting of the dot active level. We show that electron interactions in the dot diminish the characteristic for Majoranas zero bias peak arising in the transverse conductance through the dot and introduce an overall asymmetry of the conductance. They also renormalize the hybridization between the end-state Majoranas in shorter wires. The Majorana spin polarization is determined by the effective magnetic field in the dot. Phase-biased Josephson current exhibits spin polarization in thermal equilibrium, which possesses characteristic $4\pi$ periodicity, and its sign can be switched when an unpaired Majorana state is present in the junction. We also observe spin-dependent Majorana state "leaking", which can be controlled by the position of the dot level in energy scale.
\end{abstract}
\maketitle
\section{introduction}
Majorana fermions, as quantum particles, have originally been proposed by E. Majorana \cite{Majorana} in the context of particle physics. As real solutions of Dirac equation they become their own anti-particles. Recently they are predicted to appear and realized experimentally in solid state heterostructures (see recent review \cite{LutchynRev}) as complex quasiparticles (Majorana bound states-MBS) possessing properties of Majorana fermions.  Not only interesting due to their fundamental properties, they also promise the possibility of a robust to local decoherence processes quantum computation \cite{dasSarma}. The most suitable for quantum computation (Majorana braiding) \cite{Aasen} appears to be the device composed of quasi-1D semiconducting wire with large spin-orbit interaction proximitized with s-wave superconductor and subjected to an external magnetic field driving the wire into topological state, first realized experimentally by Mourik \textit{et.al.}\cite{Mourik}.

Originally, fermions proposed by Majorana possess spin one-half, recent experiments in solid state also strongly suggest that Majorana bound states exhibit spin properties, which depend on environment in which they are created. In solid state environment, they exhibit spin properties in the sense that if coupled to a non-topological object like a quantum point contact \cite{Churchill,Das,Mourik,Zhang} or a quantum dot (QD)\cite{Deng2,Deng3}, they pick a definite spin of the tunneling electron into them.  A quantum dot coupled to a topological wire is exposed to a magnetic field driving the wire into topological state. This field introduces large Zeeman field inside the dot, thus usually in the models considering such a geometry \cite{baranger,Leijnse,Lopez,Gong,Li,Campo,Schuray}, the direction of the spin of electron tunneling from/to Majorana state is assumed  to be the one of the lowest in energy QD spin sublevel. As the external magnetic field  is the largest energy scale, the dot exposed to it can be regarded as effectively non-interacting. Similarly is frequently assumed in the modelling of the Josephson junctions.\cite{Gao,Sothmann,Wang}

Recent experiments of spin polarized STM  tunneling into Majorana mode at the vortex center of a topological superconductor \cite{Sun} also show that the Majorana mode picks the spin of the electron which is parallel to the external magnetic field. It is manifested in a substantially higher MBS zero-bias peak in the differential conductance as compared to the case of the antiparallel STM tip polarization and external magnetic field. Thus, due to the spin-polarized Majorana state, equal spin Andreev reflections (ESAR) are favored in the transport \cite{He}.

Recently it has been shown theoretically \cite{klin1} that spin-dependent tunneling between a quantum dot and a topological superconducting wire can be altered with the distance of the dot from the Majorana state at the end of the wire. In the geometry of the dot placed between two sections of topological wire, as a result of the coupling to the two flanked Majorana states, a resultant magnetic field is created in the dot introducing its spin polarization.

\begin{figure} [ht]
\epsfxsize=0.45\textwidth
\epsfysize=0.45\textwidth
\epsfbox{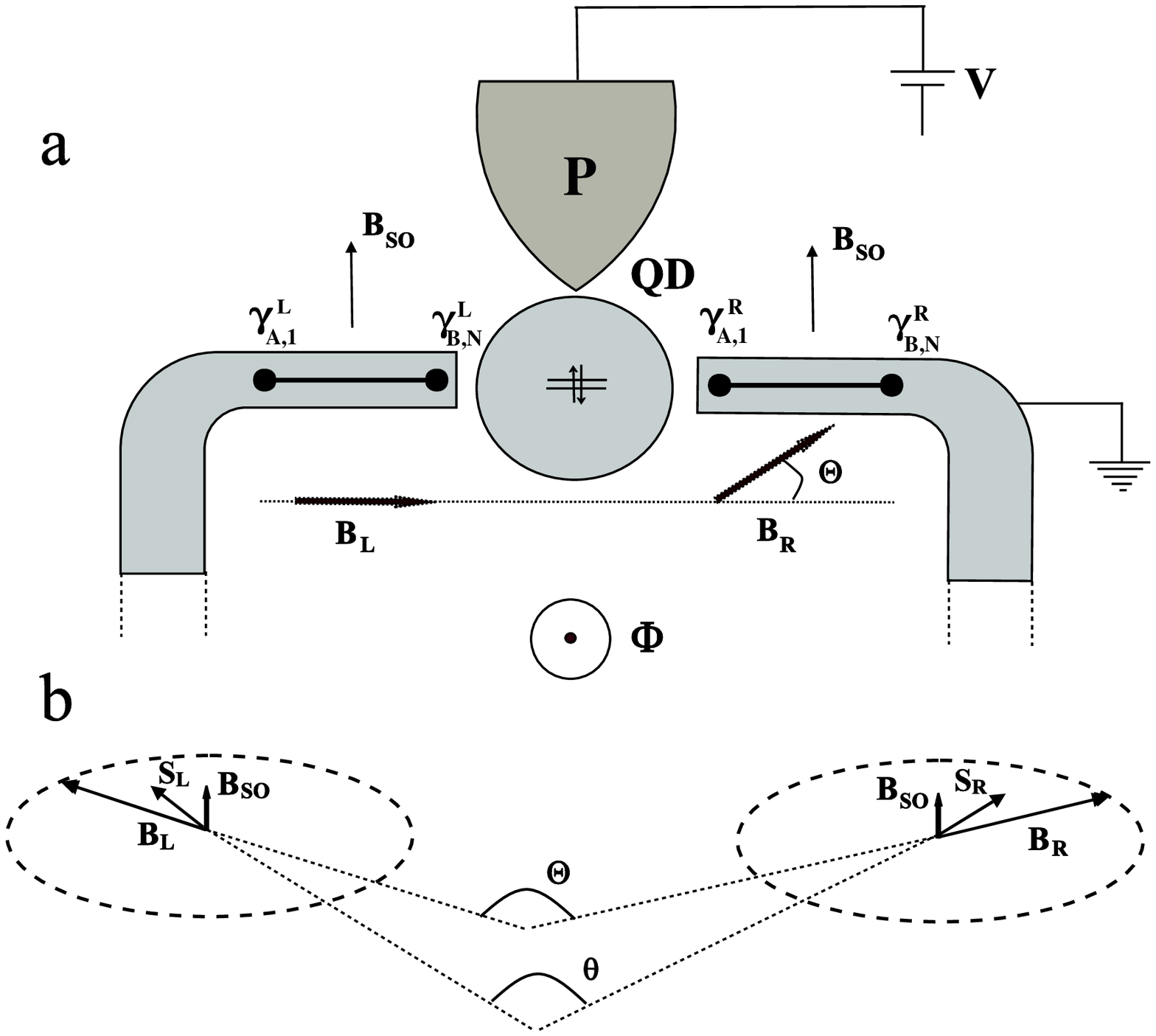}
\caption{\label{fig1}a): Schematic of the device layout showing the formation of Majorana bound states at the ends of the wires, and the mutual alignment of the  magnetic fields $\vec{B}_{L}$ and $\vec{B}_{R}$, forming an angle $\Theta$ and driving the wires into topological state. b): schematic of the rotation of the fields  $\vec{B}_{L}$ and $\vec{B}_{R}$  in the plane perpendicular to the spin-orbit field $\vec{B}_{SO}$. The direction of the effective spins, $\vec{S}_{\alpha}$, is determined by the superposition of external magnetic field and the Rashba field in each wire, and they form an angle $\theta$. }
\end{figure}

In our work we address the question, how spin dependent tunneling can be manifested in Josephson junction hosting Majorana bound states. To investigate this problem we place a  small quantum dot with one active spin-degenerate spatial level inside the junction in order to mediate the transport between quasi-1D topological wires. Contrary to the geometry of QD side-coupled  to a topological wire, in the present geometry we have the freedom of magnetic field alteration inside the dot, which allows a controllable manipulation of the MBS spin polarization. Namely, when the external magnetic fields $\vec{B}_{L}$ and $\vec{B}_{R}$, driving the wires into the topological phase, are parallel, it is natural to assume that the electrons tunneling from the dot to the nearby Majorana states in the left and the right wire possess the same spin. In this case, the dot level exposed to the effective field composed of $\vec{B}_{L}$ and $\vec{B}_{R}$, experiences large Zeeman splitting, leaving active in transport the lower (spin-down) sub-level only, and thus the dot can be regarded as effectively non-interacting. The picture changes when $\vec{B}_{L}$ and $\vec{B}_{R}$, lying in the common plane perpendicular to the spin-orbit (SO) field in the wires (see Fig.~\ref{fig1}b), are rotated to be antiparallel. In the side-coupled geometry of the dot and only the left wire, the Majorana state picks an electron of a definite spin from the dot, whereas for the dot and only the right wire, the MBS picks the electron of  spin opposite to the spin tunneling from the left wire.

In the present transmission geometry with a junction mediated by a quantum dot, we assume that the dot is made of the same materials as the wires, and the wires are strongly coupled via the dot (see the experiment by Chang \textit{et al.}\cite{Chang}), allowing the MBS adjacent to the dot to convert coherently into a Dirac fermion inside the dot. For parallel magnetic fields $\vec{B}_{L}$ and $\vec{B}_{R}$, the operator  $d$, ascribed to the dot level, has a definite spin, the same as the lower active QD spin sublevel. However, when the magnetic fields $\vec{B}_{L}$ and $\vec{B}_{R}$ are antiparallel, the $d$ operator possesses  equal admixtures of spin-up and spin-down electrons. Also the QD spatial level retains its spin degeneracy because $\vec{B}_{L}$ and $\vec{B}_{R}$ cancel each other inside the dot.

We investigate this spin-dependent feature for an angle between magnetic fields $\vec{B}_{L}$ and $\vec{B}_{R}$ ranging from zero to $\pi$. Dependent on this angle, the dot experiences variable Zeeman splitting, which allows also to address the question of the influence of electron interactions inside the dot on the Majorana states hybridizing through the junction.

Based on an effective low energy model, we derive the Hamiltonian of the Josephson junction mediated by a spinful quantum dot, in which Majorana states hybridize into the dot's bound state. Within the model, spin components of Majorana states can be separately analysed, dependent on the effective magnetic field in the dot. We showed, in agreement with recent experiments, that the MBS spin polarization is determined by the resultant magnetic field in the junction. Electron interactions inside the dot, treated in the Hubbard I approximation, diminish the height of the characteristic Majorana zero bias peak in the transverse conductance through the dot. The overall conductance peak becomes also asymmetric as a result of the change of the parity of the dot. The hybridization between end Majorana states in each wire forming the junction is also renormalized by electron interactions. We also show that the characteristic $4\pi$ periodicity of the Majorana-mediated Josephson junction vs. superconducting phase difference emerges in the thermally averaged spin polarization of the Josephson current. The conservation of parity is not required for this effect to appear. Moreover, in the presence of an unpaired Majorana bound state in the junction, the sign switching of the polarization appears with the change of the phase bias and it exhibits also $4\pi$ period. A fraction of the results has been presented in the conference proceedings.\cite{PSConf}

\section{Description of the model}
\subsection{Hamiltonian of the system}
The model schematic is depicted in Fig.~\ref{fig1}. The effective spin, $\vec{S}_{\alpha}$, in each wire is determined by the superposition of the Rashba field in the wire, $\vec{B}_{SO}$, and the external field $\vec{B}_{\alpha}$; they should be mutually perpendicular in order to generate topological state \cite{Mourik}. When $\vec{B}_{L}$ and $\vec{B}_{R}$ are antiparallel, they cancel each other in the dot, retaining its level spin degeneracy and on-site Coulomb interactions. In our model we consider the case when the fields $\vec{B}_{L}$ and $\vec{B}_{R}$ form an angle $\Theta$ and can be rotated independently in the plane perpendicular to the Rashba field, which we assume equal in both the wires. The angle $\theta$ defines the relative direction of $\vec{S}_{\alpha}$ spins in the left and right wire. Due to the presence of the strong spin-orbit interaction in the wires, the spin quantum number is no longer a conserved quantity. However, there is a tendency for absorbing or emitting electrons from/to the dot by MBS with definite spin majority in $z$-direction, defined by the external magnetic field. In this sense the effective spin $\vec{S}_{\alpha}$ is discussed in the present model.

The estimated spin-orbit energy scale in InSb wires \cite{Mourik} is $\alpha^{2}m^{*}/(2\hbar)\approx 50 \mu eV$ ($\alpha$ being the Rashba parameter, $m^{*}=0.015 m_{e}$ the electron effective mass). The ratio of Zeeman energy to applied magnetic field \cite{Mourik} is $E_{z}/B\approx 1500 \mu eV/T$, form which we can estimate the energy corresponding to the initial magnetic field \cite{Mourik}, $B_{\alpha}=0.15 T$, driving the wire into topological phase.  It yields the ratio $B_{\alpha}/B_{so}\sim 5$, which allows us to take $\Theta\approx\theta$ (see Fig.~\ref{fig1}b) and assume that the direction of the tunneling spin from the dot into adjacent Majorana states on both sides of the junction is mainly determined by external fields $\vec{B}_{\alpha}$, similarly like for Majorana  state formed in the vortex \cite{Sun}.

Due to the non-local nature of the operation of the magnetic field, for $\Theta\sim\pi$ it could arise a configuration that some parts of the wires adjacent to the dot were driven out of the topological state. Supported by the recent experiment \cite{Deng3}, we assume that there is a "leaking" of Majorana states into that region and in such a way they are able to hybridize inside the dot.

The quantum dot placed inside the Josephson junction is exposed to the action of the magnetic field $\overrightarrow{B}_{tot}=\overrightarrow{B}_{L}+\overrightarrow{B}_{R}$. In further considerations we assume for simplicity $|\overrightarrow{B}_{L}|=|\overrightarrow{B}_{R}|\equiv B$. The Zeeman field, which lifts the spin degeneracy of the dot spatial level, is defined as  $E_{z}=|g|\mu_{B}B_{tot}(\Theta)/2$ with $B_{tot}(\Theta)=2 B\cos(\Theta/2)$. It yields $E_{z}=E_{z}^{0}\cos(\Theta/2)$ with $E_{z}^{0}=|g|\mu_{B}B$. Under the action of the Zeeman field we obtain $\epsilon_{\uparrow/\downarrow}=\epsilon_{d}\pm E_{z}^{0}\cos(\Theta/2)-V_{g}$. The energy structure of the dot can be shifted by an underlying gate voltage, $V_{g}$, with respect to Fermi level, $\epsilon_{F}=0$, and we assume $\epsilon_{d}=\epsilon_{F}$ for $V_{g}=0$.
The dot Hamiltonian is of the form:
\begin{equation}
\label{Hdot}
H_{QD}=\sum_{\sigma=\uparrow,\downarrow}\epsilon_{\sigma}d^{\dagger}_{\sigma}d_{\sigma}+U n_{\downarrow}n_{\uparrow}.
\end{equation}
In the following we will utilize the induced superconducting gap $\Delta$ as the energy unit.

The induced superconducting gap is $\Delta\approx 250 \mu eV$ and the wire enters the topological phase \cite{Mourik} for $B_{init}\sim 0.15 T$. With the ratio $E_{z}/B=1.5 meV/T$ this yields the initial $E_{z}\sim\Delta$. In the following we assume $E_{z}^{0}=2\Delta$ for numerical calculations.

The density of states of the dot is probed by a normal tunneling electrode; in the experiment \cite{Chang} it is made of Au and positioned above the dot. This electrode is also exposed to the field $B_{tot}$, which polarizes the electrode. This polarization is described by spin-dependent broadenings $\Gamma_{\sigma}$ ($\sigma=\downarrow, \uparrow$), which the dot level acquires by the coupling to the electrode. We model them by $\Gamma_{p\downarrow}=\Gamma_{p}\cos^{2}(\Theta/4)$ and $\Gamma_{p\uparrow}=\Gamma_{p}\sin^{2}(\Theta/4)$, where we assume $\Gamma_{p}=0.02$. For $\Theta=0$, when the electrode is fully spin-down polarized, we introduced small broadening $\Gamma_{\uparrow}=10^{-6}$ in order to retain selfconsistency of numerical calculations of the dot occupancies.

The Hamiltonian of the normal tunneling probe, to which the dot is coupled, reads:
\begin{equation}
\label{H_leads}
H_{probe}=\sum_{k,\sigma}\epsilon_{k,\sigma}c^{\dagger}_{k,\sigma}c_{k,\sigma},
\end{equation}
where $c_{k,\sigma}$ is the electron operator with momentum $k$ of the tunneling electrode with the corresponding energy $\epsilon_{k,\sigma}$.

The coupling of the tunneling electrode and  the quantum dot is described by  the Hamiltonian:
\begin{equation}
\label{H_QD_leads}
H_{QD-probe}=\sum_{k,\sigma}(t_{p}c^{\dagger}_{k,\sigma}d_{\sigma}+h.c.),
\end{equation}
$t_{p}$ being the tunneling strength between the dot and the tunneling probe, which has been assumed independent of momentum and spin.

The topological superconducting wires (TSW) are modelled by Kitaev chains \cite{Kitaev} describing a spinless p-wave superconductor. For the wire $\alpha=L,R$:
\begin{eqnarray}\nonumber
H_{\alpha}=-\mu_{\alpha}\sum_{i=1}^{N}c^{\dagger}_{i\alpha}c_{i\alpha}-\\
\sum_{i=1}^{N-1}(t_{\alpha}c^{\dagger}_{i\alpha}c_{i+1\alpha}
+|\Delta_{\alpha}|e^{i\phi_{\alpha}}c_{i\alpha}c_{i+1\alpha}+h.c.),
\end{eqnarray}
where $c_{i\alpha}$ is a spinless fermion operator for the site $i$, and $\mu_{\alpha}$, $t_{\alpha}$ and $|\Delta_{\alpha}|e^{i\phi_{\alpha}}$ are the chemical potential, tunneling matrix element and the superconducting order parameter, respectively. In the simplest topological phase for $\mu_{\alpha}=0$ and $|\Delta_{\alpha}|=t_{\alpha}\equiv t$, the Hamiltonian of the wires assumes the form:
\begin{equation}
H_{\alpha}=-t\sum_{i=1}^{N-1}(c^{\dagger}_{i\alpha}c_{i+1\alpha}+e^{i\phi_{\alpha}}c_{i\alpha}c_{i+1\alpha}+h.c.)
\end{equation}

Dirac fermions in the Kitaev chains can be expressed in terms of two Majorana fermions. For each $i$-site of the wire $\alpha$:
\begin{equation}
\label{c_fermion}
c_{i\alpha}=e^{-i(\phi_{\alpha}/2)}(1/2)(\gamma^{\alpha}_{B,i}+i\gamma^{\alpha}_{A,i}),
\end{equation}
where $\gamma^{\alpha}_{i}=(\gamma^{\alpha}_{i})^{\dagger}$ are Majorana operators satisfying $\{\gamma_{i}^{\alpha},\gamma_{j}^{\alpha'}\}=2\delta_{\alpha,\alpha'}\delta_{i,j}$.  The TSW Hamiltonian in this notation describes the coupling between the MBS of the adjacent sites:
\begin{equation}
H_{\alpha}=-it\sum_{i=1}^{N-1}\gamma^{\alpha}_{B,i}\gamma^{\alpha}_{A,i+1},
\end{equation}
and two end-site MBS $\gamma^{\alpha}_{A,1}$ and $\gamma^{\alpha}_{B,N}$ in the wire $\alpha$ remain unpaired.
Upon refermionization by the introduction of Dirac fermions composed of Majorana fermions of adjacent sites:
\begin{equation}
\label{f_fermion}
 f_{\alpha i}=(1/2)(\gamma^{\alpha}_{A,i+1}+i\gamma^{\alpha}_{B,i}),
\end{equation}
 the wire Hamiltonian becomes diagonal:
\begin{equation}
\label{H_wire_fer}
H_{\alpha}=t\sum_{i=1}^{N-1}(2f^{\dagger}_{\alpha i}f_{\alpha i}-1).
\end{equation}

The dot level $\epsilon_{d}$ is coherently coupled to the end-wire MBS: $\gamma^{L}_{B,N}$ and $\gamma^{R}_{A,1}$, which are converted into the Dirac fermion $d$ inside the dot. Thus, all the information, in particular on the superconducting phase difference, is encoded in the Dirac fermion $d$ in the dot:
\begin{equation}
\label{d_fermion}
d=\frac{1}{2}(\gamma^{R}_{A,1}+i\gamma^{L}_{B,N}).
\end{equation}
Dependent on the angle $\Theta$ formed by the magnetic fields $\vec{B}_{L}$ and $\vec{B}_{R}$ the Majorana states $\gamma^{L}_{B,N}$ and $\gamma^{R}_{A,1}$ can pick electrons with different spin orientations tunneling from the dot, see Fig.~\ref{fig1}. This is described by the superposition of spin-down and spin-up operators for the QD operator $d$:
\begin{equation}
\label{d_composite}
d=\cos\left(\frac{\Theta}{4}\right)d_{\downarrow}+\sin\left(\frac{\Theta}{4}\right)d_{\uparrow}.
\end{equation}

The Majorana states  calculated from Eqs.~(\ref{d_fermion}) and (\ref{d_composite}) uncover their spin-dependent components, which themselves possess the properties of Majorana operators:
\begin{widetext}
\begin{eqnarray}
\label{Majorana_s1}
\gamma^{L}_{B,N}=i\cos\left(\frac{\Theta}{4}\right)(d^{\dagger}_{\downarrow}-d_{\downarrow})+i\sin\left(\frac{\Theta}{4}\right)(d^{\dagger}_{\uparrow}-d_{\uparrow})=
\gamma^{L}_{B,N,\downarrow}+\gamma^{L}_{B,N,\uparrow}\\
\label{Majorana_s2}
\gamma^{R}_{A,1}=\cos\left(\frac{\Theta}{4}\right)(d^{\dagger}_{\downarrow}+d_{\downarrow})+
\sin\left(\frac{\Theta}{4}\right)(d^{\dagger}_{\uparrow}+d_{\uparrow})=\gamma^{R}_{A,1,\downarrow}+\gamma^{R}_{A,1,\uparrow}.
\end{eqnarray}
\end{widetext}

The specific geometry of the model allows to control the amounts of these spin components.  For $\Theta=0$ the spinless Kitaev model is recovered (with suppressed spin index)\cite{Leijnse2}. For $\Theta=\pi$ the QD level is spin-degenerate and there are equal contributions from both spin sectors, thus no specific direction of the spin can be distinguished. The coherent conversion of MBS into Dirac dot's fermion in this case, resembles such a conversion of Majorana states of opposite chiralities propagating along a magnetic domain on the surface of topological insulator, proximitized to superconductor \cite{Akhmerov}.

For $\Theta<\pi$, however, the Majorana states acquire finite spin polarizations. Moreover, a finite Zeeman splitting in the dot allows to distinguish between the Majorana spin components, which is reflected in the transverse conductance through the dot, as presented below.

In order to properly describe the superconducting phase evolution across the junction mediated by the dot, one has to take into account the proximity effect of the superconducting wires on the dot. An induced superconducting order is created inside the dot in the vicinity of the left and the right wire. In the low energy limit of our model, $\omega \ll\Delta$, the Hamiltonian describing an induction of superconductivity inside the dot near the wire $\alpha$ can be written \cite{Martin-Rodero}:
\begin{equation}
H^{prox}_{\alpha}=|\Delta_{\alpha}^{ind}|e^{\phi_{\alpha}}dd+h.c.,
\end{equation}
where $\Delta_{\alpha}^{ind}$ is the induced superconducting order parameter. Thus, a quasiparticle with phase $\phi_{\alpha}$ tunneling from the wire $\alpha$ into the dot, experiences the proximity of the $\alpha'$ wire through the phase $\phi_{\alpha'}$ induced inside the dot. To describe this effect we define a phase-dependent QD fermion $d_{\alpha}=\exp(-i\phi_{\alpha}/2)d$ and write the tunneling Hamiltonian in the form:
\begin{equation}
H_{tun}=-t_{L}(c_{L,N}^{\dagger}d_{R}+h.c.)-t_{R}(c_{R,1}^{\dagger}d_{L}+h.c.),
\end{equation}
where $t_{\alpha}$ is the tunneling strength between the wire $\alpha$ and the dot.
Taking into account Eq.~(\ref{c_fermion}), the tunneling Hamiltonian can be written in terms of Majorana states revealing its dependencies on the superconducting phase difference:
\begin{widetext}
\begin{eqnarray}
\label{H_tun}
\nonumber
H_{tun}=-\frac{i}{2}t_{L}\left[\cos(\frac{\Delta\phi}{2})(\gamma^{L}_{A,N}\gamma^{L}_{B,N}-\gamma^{L}_{B,N}\gamma^{R}_{A,1})
-\sin(\frac{\Delta\phi}{2})\gamma^{L}_{A,N}\gamma^{R}_{A,1}\right]\\
-\frac{i}{2}t_{R}\left[\cos(\frac{\Delta\phi}{2})(\gamma^{R}_{A,1}\gamma^{L}_{B,N}-\gamma^{R}_{B,1}\gamma^{R}_{A,1})
+\sin(\frac{\Delta\phi}{2})\gamma^{R}_{B,1}\gamma^{L}_{B,N}\right],
\end{eqnarray}
\end{widetext}
where, without the loss of generality, we introduced $\phi_{R}=-\phi_{L}=\Delta\phi/2$. Retaining the essential physics, in further analysis we confine ourselves to the simpler version of Eq.~(\ref{H_tun}) by taking into account the hybridization only of the Majoranas adjacent to the dot:

\begin{equation}
\label{H_tun_simp}
H_{tun}=-\frac{i}{2}(t_{L}+t_{R})\cos(\frac{\Delta\phi}{2})\gamma^{L}_{B,N}\gamma^{R}_{A,1}.
\end{equation}
Then, with the help Eqs.~(\ref{Majorana_s1}) and (\ref{Majorana_s2}), we refermionize Eq.~(\ref{H_tun_simp}):
\begin{widetext}
\begin{equation}
\label{H_tun_simp_fer}
H_{tun}=-(t_{L}+t_{R})\cos(\frac{\Delta\phi}{2})\left[\cos^{2}(\frac{\Theta}{4})(d^{\dagger}_{\downarrow}d_{\downarrow}-\frac{1}{2})+
\sin^{2}(\frac{\Theta}{4})(d^{\dagger}_{\uparrow}d_{\uparrow}-\frac{1}{2})+\sin(\frac{\Theta}{4})\cos(\frac{\Theta}{4})(d^{\dagger}_{\uparrow}d_{\downarrow}+h.c.) \right]
\end{equation}
\end{widetext}
One notes that in the refermionized Hamiltonian  the terms describing spin-flip processes inside the dot appear  naturally. Such processes  prevent the appearance of the spin blockade in the tunneling through the junction, which could arise for anti-parallel magnetic fields in the wires.

Lastly, we introduce the Hamiltonian describing a possible finite hybridization between the end-Majoranas within a given wire. It is described by the overlap  of the Majorana wave functions, $\epsilon_{\alpha}\sim e^{-L_{\alpha}/\xi}$, where $\xi$ is the induced superconducting coherence length and $L$ - the wire length. It has been proven experimentally, that for the lengths $L_{\alpha}\sim$ 500 $nm$ \cite{Mourik} this hybridization was negligible, and the exponential decay with the increase of the wire length has also been demonstrated \cite{Albrecht}. Taking into account the characteristic spin dependence of the MBS adjacent to the dot, described by Eqs.~(\ref{Majorana_s1})and (\ref{Majorana_s2}), the hybridization of the MBS located at the ends of the wire $\alpha$ can also be written for each spin component:

\begin{equation}
\label{H_hyb}
H_{hyb}=-\sum_{\alpha=L,R}\sum_{\sigma=\downarrow,\uparrow} i\frac{\epsilon_{\alpha}}{2}\gamma^{\alpha}_{A,1,\sigma}\gamma^{\alpha}_{B,N\sigma}.
\end{equation}

As a result, a non-local Dirac fermion can be created out of two MBS in each wire with spin components:
\begin{equation}
\label{f_alpha}
f_{\alpha,\sigma}=(1/2)(\gamma^{\alpha}_{A,1,\sigma}+i \gamma^{\alpha}_{B,N,\sigma}).
\end{equation}

In our model this fermion is tracked by a spin index, introduced by the possibility of altering the magnetic field direction in the dot region.
Using Eq.~(\ref{f_alpha}) to obtain the MBS farther from the dot, $\gamma^{L}_{A,1,\sigma}$ and $\gamma^{R}_{B,N,\sigma}$,  and Eqs.~(\ref{Majorana_s1}), (\ref{Majorana_s2}) for the Majoranas close to the dot, $\gamma^{L}_{B,N,\sigma}$ and $\gamma^{R}_{A,1,\sigma}$, $H_{hyb}$ can be written in terms of Dirac fermions:
\begin{widetext}
\begin{eqnarray}
\label{H_ov_fer}\nonumber
H_{hyb}=\frac{\epsilon_{L}}{2}\left[\cos\left(\frac{\Theta}{4}\right)(f^{\dagger}_{L\downarrow}d^{\dagger}_{\downarrow}-f^{\dagger}_{L\downarrow}d_{\downarrow}+h.c.)+
\sin\left(\frac{\Theta}{4}\right)(f^{\dagger}_{L\uparrow}d^{\dagger}_{\uparrow}-f^{\dagger}_{L\uparrow}d_{\uparrow}+h.c.) \right]-\\
\frac{\epsilon_{R}}{2}\left[\cos\left(\frac{\Theta}{4}\right)(f^{\dagger}_{R\downarrow}d^{\dagger}_{\downarrow}+f^{\dagger}_{R\downarrow}d_{\downarrow}+h.c.)+
\sin\left(\frac{\Theta}{4}\right)(f^{\dagger}_{R\uparrow}d^{\dagger}_{\uparrow}+f^{\dagger}_{R\uparrow}d_{\uparrow}+h.c.) \right].
\end{eqnarray}
\end{widetext}
The above Hamiltonian, Eq.~(\ref{H_ov_fer}), can also be formally recognized as two- and one-particle tunneling with the strength $\epsilon_{\alpha}/2$ between the dot and the extended fermionic state made of hybridized Majorana states in the $\alpha$ wire. One notes that for $\Theta=0$ the limit of equal spin Andreev reflections (ESAR) is recovered, recently discussed in the context of tunneling between a $1D$-topological wire and a non-topological electrode \cite{He}.

\subsection{Green's function of the dot}
In order to describe basic characteristics of the junction, we calculate the QD retarded Green's function, $\langle\langle d;d^{\dagger}\rangle\rangle=-i\theta(t-t')\langle d(t)d^{\dagger}(t')+d^{\dagger}(t')d(t) \rangle$, in presence of the tunneling electrode and the quantum wires. Taking into account Eq.~(\ref{d_composite}), we approximate:
\begin{equation}
\label{composite_GF}
\langle\langle d;d^{\dagger}\rangle\rangle\cong\cos^{2}(\frac{\Theta}{4})\langle\langle d_{\downarrow};d^{\dagger}_{\downarrow}\rangle\rangle+
\sin^{2}(\frac{\Theta}{4})\langle\langle d_{\uparrow};d^{\dagger}_{\uparrow}\rangle\rangle,
\end{equation}
where we have neglected the terms describing spin "mixing". However, the Green's functions $\langle\langle d_{\downarrow};d^{\dagger}_{\downarrow}\rangle\rangle$ and $\langle\langle d_{\uparrow};d^{\dagger}_{\uparrow}\rangle\rangle$ are calculated in the presence of all  spin "mixing" terms in the Hamiltonian, electron interactions and the mixing in the tunneling Hamiltonian. Namely, the spin-down component  $\langle\langle d_{\downarrow};d^{\dagger}_{\downarrow}\rangle\rangle$ is calculated
within the equation of motion approach, taking into account Eqs.~(\ref{Hdot}), (\ref{H_leads}), (\ref{H_QD_leads}), (\ref{H_tun_simp_fer}) and  (\ref{H_ov_fer}). We obtain a set of equations in frequency domain:
\begin{widetext}
\begin{eqnarray}
\nonumber
(\omega-\epsilon_{\downarrow}+t\alpha^2+i\frac{\Gamma_{p\downarrow}}{2})\langle\langle d_{\downarrow};d^{\dagger}_{\downarrow}\rangle\rangle=1-\\
t\alpha\beta\langle\langle d_{\uparrow};d^{\dagger}_{\downarrow}\rangle\rangle-\alpha\{\frac{\epsilon_{L}}{2}[\langle\langle f^{\dagger}_{L\downarrow};d^{\dagger}_{\downarrow}\rangle\rangle+\langle\langle f_{L\downarrow};d^{\dagger}_{\downarrow}\rangle\rangle]-
\frac{\epsilon_{R}}{2}[\langle\langle f^{\dagger}_{R\downarrow};d^{\dagger}_{\downarrow}\rangle\rangle-\langle\langle f_{R\downarrow};d^{\dagger}_{\downarrow}\rangle\rangle]\}+U\langle\langle n_{\uparrow}d_{\downarrow};d^{\dagger}_{\downarrow}\rangle\rangle,\\
\label{elim1}
\omega\langle\langle f^{\dagger}_{L\downarrow};d^{\dagger}_{\downarrow}\rangle\rangle=
\alpha\frac{\epsilon_{L}}{2}[\langle\langle d^{\dagger}_{\downarrow};d^{\dagger}_{\downarrow}\rangle\rangle-\langle\langle d_{\downarrow};d^{\dagger}_{\downarrow}\rangle\rangle],\\
\label{elim2}
\langle\langle f_{L\downarrow};d^{\dagger}_{\downarrow}\rangle\rangle=\langle\langle f^{\dagger}_{L\downarrow};d^{\dagger}_{\downarrow}\rangle\rangle,\\
\label{elim3}
\omega\langle\langle f^{\dagger}_{R\downarrow};d^{\dagger}_{\downarrow}\rangle\rangle=
\alpha\frac{\epsilon_{R}}{2}[\langle\langle d^{\dagger}_{\downarrow};d^{\dagger}_{\downarrow}\rangle\rangle+\langle\langle d_{\downarrow};d^{\dagger}_{\downarrow}\rangle\rangle],\\
\label{elim4}
\langle\langle f_{R\downarrow};d^{\dagger}_{\downarrow}\rangle\rangle=-\langle\langle f^{\dagger}_{R\downarrow};d^{\dagger}_{\downarrow}\rangle\rangle,\\
\nonumber
(\omega+\epsilon_{\downarrow}-t\alpha^2+i\frac{\Gamma_{p\downarrow}}{2})\langle\langle d^{\dagger}_{\downarrow};d^{\dagger}_{\downarrow}\rangle\rangle=
t\alpha\beta\langle\langle d^{\dagger}_{\uparrow};d^{\dagger}_{\downarrow}\rangle\rangle+\\
\alpha\{\frac{\epsilon_{L}}{2}[\langle\langle f^{\dagger}_{L\downarrow};d^{\dagger}_{\downarrow}\rangle\rangle+\langle\langle f_{L\downarrow};d^{\dagger}_{\downarrow}\rangle\rangle]+
\frac{\epsilon_{R}}{2}[\langle\langle f^{\dagger}_{R\downarrow};d^{\dagger}_{\downarrow}\rangle\rangle-\langle\langle f_{R\downarrow};d^{\dagger}_{\downarrow}\rangle\rangle]\}-U\langle\langle n_{\uparrow}d^{\dagger}_{\downarrow};d^{\dagger}_{\downarrow}\rangle\rangle,\\
\nonumber
(\omega-\epsilon_{\uparrow}+t\beta^{2}+i\frac{\Gamma_{p\uparrow}}{2})\langle\langle d_{\uparrow};d^{\dagger}_{\downarrow}\rangle\rangle=
-t\alpha\beta\langle\langle d_{\downarrow};d^{\dagger}_{\downarrow}\rangle\rangle-\\
\beta\{\frac{\epsilon_{L}}{2}[\langle\langle f^{\dagger}_{L\downarrow};d^{\dagger}_{\downarrow}\rangle\rangle+\langle\langle f_{L\downarrow};d^{\dagger}_{\downarrow}\rangle\rangle]-\frac{\epsilon_{R}}{2}[\langle\langle f^{\dagger}_{R\downarrow};d^{\dagger}_{\downarrow}\rangle\rangle-\langle\langle f_{R\downarrow};d^{\dagger}_{\downarrow}\rangle\rangle]\}+U\langle\langle n_{\downarrow}d_{\uparrow};d^{\dagger}_{\downarrow}\rangle\rangle,\\
\nonumber
(\omega+\epsilon_{\uparrow}-t\beta^{2}+i\frac{\Gamma_{p\uparrow}}{2})\langle\langle d^{\dagger}_{\uparrow};d^{\dagger}_{\downarrow}\rangle\rangle=
t\alpha\beta\langle\langle d^{\dagger}_{\downarrow};d^{\dagger}_{\downarrow}\rangle\rangle+\\
\beta\{\frac{\epsilon_{L}}{2}[\langle\langle f^{\dagger}_{L\downarrow};d^{\dagger}_{\downarrow}\rangle\rangle+\langle\langle f_{L\downarrow};d^{\dagger}_{\downarrow}\rangle\rangle]+\frac{\epsilon_{R}}{2}[\langle\langle f^{\dagger}_{R\downarrow};d^{\dagger}_{\downarrow}\rangle\rangle-\langle\langle f_{R\downarrow};d^{\dagger}_{\downarrow}\rangle\rangle]\}
-U\langle\langle n_{\downarrow}d^{\dagger}_{\uparrow};d^{\dagger}_{\downarrow}\rangle\rangle.
\end{eqnarray}
\end{widetext}
Note that the level broadenings $i(\Gamma_{p\sigma}/2)$ introduced by the corresponding Green's functions $\pm\sum_{k}t_{p}\langle\langle c^{(\dagger)}_{k\sigma};d^{\dagger}_{\downarrow}\rangle\rangle$ have been already incorporated  in the above equations.

The density of states in the probe electrode has been treated in the wide band approximation, yielding $\Gamma_{p\sigma}=2\pi|t_{p}|^{2}\rho_{p\sigma}$, where $\rho_{p\sigma}$ is featureless density of states of the electrode.

After eliminating $\langle\langle f_{\alpha\downarrow};d^{\dagger}_{\downarrow}\rangle\rangle$ and $\langle\langle f^{\dagger}_{\alpha\downarrow};d^{\dagger}_{\downarrow}\rangle\rangle$ ($\alpha=L, R$) Green's functions, we obtain a more compact set of equations:
\begin{widetext}
\begin{eqnarray}
\label{eom1}
(\omega-\epsilon_{\downarrow-}+i\frac{\Gamma_{p\downarrow}}{2})\langle\langle d_{\downarrow};d^{\dagger}_{\downarrow}\rangle\rangle=1-
\alpha^{2}\epsilon_{-}\langle\langle d^{\dagger}_{\downarrow};d^{\dagger}_{\downarrow}\rangle\rangle-t\alpha\beta\langle\langle d_{\uparrow};d^{\dagger}_{\downarrow}\rangle\rangle+U\langle\langle n_{\uparrow}d_{\downarrow};d^{\dagger}_{\downarrow}\rangle\rangle,\\
\label{eom2}
(\omega+\epsilon_{\downarrow+}+i\frac{\Gamma_{p\downarrow}}{2})\langle\langle d^{\dagger}_{\downarrow};d^{\dagger}_{\downarrow}\rangle\rangle=-
\alpha^{2}\epsilon_{-}\langle\langle d_{\downarrow};d^{\dagger}_{\downarrow}\rangle\rangle+t\alpha\beta\langle\langle d^{\dagger}_{\uparrow};d^{\dagger}_{\downarrow}\rangle\rangle-U\langle\langle n_{\uparrow}d^{\dagger}_{\downarrow};d^{\dagger}_{\downarrow}\rangle\rangle,\\
\label{eom3}
(\omega-\epsilon_{\uparrow-}+i\frac{\Gamma_{p\uparrow}}{2})\langle\langle d_{\uparrow};d^{\dagger}_{\downarrow}\rangle\rangle=-
\beta^{2}\epsilon_{-}\langle\langle d^{\dagger}_{\uparrow};d^{\dagger}_{\downarrow}\rangle\rangle-t\alpha\beta\langle\langle d_{\downarrow};d^{\dagger}_{\downarrow}\rangle\rangle+U\langle\langle n_{\downarrow}d_{\uparrow};d^{\dagger}_{\downarrow}\rangle\rangle,\\
\label{eom4}
(\omega+\epsilon_{\uparrow+}+i\frac{\Gamma_{p\uparrow}}{2})\langle\langle d^{\dagger}_{\uparrow};d^{\dagger}_{\downarrow}\rangle\rangle=-
\beta^{2}\epsilon_{-}\langle\langle d_{\uparrow};d^{\dagger}_{\downarrow}\rangle\rangle+t\alpha\beta\langle\langle d^{\dagger}_{\downarrow};d^{\dagger}_{\downarrow}\rangle\rangle-U\langle\langle n_{\downarrow}d^{\dagger}_{\uparrow};d^{\dagger}_{\downarrow}\rangle\rangle,
\end{eqnarray}
\end{widetext}
where we have introduced the following notation: $\alpha\equiv\cos(\Theta/4)$, $\beta\equiv\sin(\Theta/4)$, $t\equiv (t_{L}+t_{R})\cos(\Delta\phi/2)$,
$\epsilon_{\mp}\equiv(\epsilon_{L}^{2}\mp\epsilon_{R}^{2})/2\omega$ and
$\epsilon_{\downarrow\mp}\equiv\epsilon_{\downarrow}-\alpha^{2}(t\mp \epsilon_{+})$, $\epsilon_{\uparrow\mp}\equiv\epsilon_{\uparrow}-\beta^{2}(t\mp \epsilon_{+})$. We also introduce renormalized dot's spin sublevels: $\tilde{\epsilon}_{\downarrow}\equiv\epsilon_{\downarrow}-t\alpha^{2}$ and  $\tilde{\epsilon}_{\uparrow}\equiv\epsilon_{\uparrow}-t\beta^{2}$.

At this stage Eqs.(\ref{eom1})-(\ref{eom4}) are exact, and we need some approximations for calculations of the effects of electron correlations hidden in the Green's functions with $U$ prefactor. If we decoupled the above Green's functions in the Hartree-Fock manner: $\langle\langle n_{\sigma}d^{(\dagger)}_{\sigma'};d^{\dagger}_{\downarrow}\rangle\rangle
\cong\langle n_{\sigma}\rangle \langle\langle d^{(\dagger)}_{\sigma'};d^{\dagger}_{\downarrow}\rangle\rangle$, we would obtain just the renormalizations of the dot levels $\epsilon_{\sigma}$ by the terms $\mp\langle n_{\sigma'}\rangle U$. Instead, dictated by the underlying physics of the model,  we neglect the processes which are less probable.

First we neglect the Green's functions $\langle\langle n_{\sigma}d^{\dagger}_{\sigma'};d^{\dagger}_{\downarrow}\rangle\rangle$ ($\sigma,\sigma'=\downarrow, \uparrow$) in Eqs.~(\ref{eom2}) and (\ref{eom4}), which describe creation of two electrons on the dot level, which already has  some occupation probability. Moreover, since the dot is only weakly coupled to the tunneling electrode, we neglect the contribution to electron interactions originating from spin-flip processes, associated with the electron tunneling from/to the electrode, which would lead to the formation of the Kondo resonance. From the experimental point of view, this effect is not desirable as it would obscure Majorana states formation at Fermi energy, making their detection more difficult. We also neglect $\langle\langle n_{\downarrow}d_{\uparrow};d^{\dagger}_{\downarrow}\rangle\rangle$ in Eq.~(\ref{eom3}), which describes spin-flip on the dot level with its finite occupancy. Under these simplifying assumptions, the only Green's function left to be evaluated is $\langle\langle n_{\uparrow}d_{\downarrow};d^{\dagger}_{\downarrow}\rangle\rangle$ in Eq.~(\ref{eom1}). After eliminating $\langle\langle f_{\alpha\downarrow};d^{\dagger}_{\downarrow}\rangle\rangle$ and $\langle\langle f^{\dagger}_{\alpha\downarrow};d^{\dagger}_{\downarrow}\rangle\rangle$ ($\alpha=L, R$) Green's functions by the use of Eqs.~(\ref{elim1}), (\ref{elim2}), (\ref{elim3}) and (\ref{elim4}), the desired equation assumes the following form:
\begin{widetext}
\begin{equation}
\label{eom5}
(\omega-\epsilon_{\downarrow}+t\alpha^{2}+i\frac{\Gamma_{p\downarrow}}{2})\langle\langle n_{\uparrow}d_{\downarrow};d^{\dagger}_{\downarrow}\rangle\rangle=
\langle n_{\uparrow}\rangle-\alpha^{2}\langle n_{\uparrow}\rangle[\epsilon_{-}\langle\langle d^{\dagger}_{\downarrow};d^{\dagger}_{\downarrow}\rangle\rangle-
\epsilon_{+}\langle\langle d_{\downarrow};d^{\dagger}_{\downarrow}\rangle\rangle],
\end{equation}
\end{widetext}
upon neglecting higher order spin-flip processes on the dot: i) originating from the coupling to the probe electrode $\langle\langle c^{\dagger}_{k\uparrow}d_{\downarrow}d_{\uparrow};d^{\dagger}_{\downarrow}\rangle\rangle$ and $\langle\langle c_{k\uparrow}d^{\dagger}_{\downarrow}d_{\uparrow};d^{\dagger}_{\downarrow}\rangle\rangle$, ii) originating from the coupling to the $\alpha$-wire:
$\langle\langle d^{\dagger}_{\uparrow}d_{\downarrow}f_{\alpha\uparrow};d^{\dagger}_{\downarrow}\rangle\rangle$, $\langle\langle d^{\dagger}_{\uparrow}d_{\downarrow}f^{\dagger}_{\alpha\uparrow};d^{\dagger}_{\downarrow}\rangle\rangle$,
$\langle\langle d_{\uparrow}d_{\downarrow}f_{\alpha\uparrow};d^{\dagger}_{\downarrow}\rangle\rangle$ and
$\langle\langle d_{\uparrow}d_{\downarrow}f^{\dagger}_{\alpha\uparrow};d^{\dagger}_{\downarrow}\rangle\rangle$.

The Green's functions describing on-site electron interactions of the electrons tunneling from the $\alpha$-wire have been decoupled in the Hartree-Fock approximation: $\langle\langle n_{\uparrow}f_{\alpha\downarrow};d^{\dagger}_{\downarrow}\rangle\rangle\cong \langle n_{\uparrow}\rangle \langle\langle f_{\alpha\downarrow};d^{\dagger}_{\downarrow}\rangle\rangle$ and $\langle\langle n_{\uparrow}f^{\dagger}_{\alpha\downarrow};d^{\dagger}_{\downarrow}\rangle\rangle\cong \langle n_{\uparrow}\rangle \langle\langle f^{\dagger}_{\alpha\downarrow};d^{\dagger}_{\downarrow}\rangle\rangle$.

Finally, the Green's function $\langle\langle n_{\uparrow}c_{k\downarrow};d^{\dagger}_{\downarrow}\rangle\rangle$ under the above assumptions introduces only level broadening, $\Gamma_{p\downarrow}$, incorporated into Eq.~(\ref{eom5}).

Taking into account Eqs.~(\ref{eom1})-(\ref{eom5}), we derive the equation for $G_{\downarrow}(\omega)\equiv\langle\langle d_{\downarrow};d^{\dagger}_{\downarrow}\rangle\rangle$:
\begin{widetext}
\begin{eqnarray}
\label{GF_down}
G_{\downarrow}(\omega)=\frac{A_{H\downarrow}}{\omega-\epsilon_{\downarrow}+\alpha^{2}(t-\epsilon_{+}A_{H\downarrow})+i\frac{\Gamma_{p\downarrow}}{2}-
\Sigma^{diag}_{\downarrow}(\omega)-\Sigma^{mix}_{\downarrow}(\omega)},\\
\nonumber
\Sigma^{diag}_{\downarrow}(\omega)=\frac{\alpha^{4}\epsilon_{-}^{2}A_{H\downarrow}}{\omega+\epsilon_{\downarrow}-\alpha^{2}(t+\epsilon_{+})+i\frac{\Gamma_{p\downarrow}}{2}},\\
\nonumber
\Sigma^{mix}_{\downarrow}(\omega)=\frac{\alpha^{2}\beta^{2}t^{2}}{\omega-\epsilon_{\uparrow}+\beta^{2}(t-\epsilon_{+})+i\frac{\Gamma_{p\uparrow}}{2}},\\
\nonumber
A_{H\downarrow}=\frac{\omega-\epsilon_{\downarrow}-U(1-\langle n_{\uparrow}\rangle)+\alpha^{2}t+i\frac{\Gamma_{p\downarrow}}{2}}
{\omega-\epsilon_{\downarrow}-U+\alpha^{2}t+i\frac{\Gamma_{p\downarrow}}{2}}.
\end{eqnarray}
\end{widetext}
In the derivation we retained only the first terms $\Sigma^{diag}_{\downarrow}$ and $\Sigma^{mix}_{\downarrow}$ of the selfenergy, diagonal in spin indices and mixing the spin indices, respectively. The corresponding spin-up Green's function component, $G_{\uparrow}$, can be obtained from Eq.~(\ref{GF_down}) by the exchange of spin indices
$\downarrow\rightleftarrows\uparrow$ \textit{and} $\alpha\rightleftarrows\beta$.

It is worth noticing that for $U=0$ one obtains $A_{H\sigma}=1$, and the QD Green's function assumes the form as the one from derivation for non-interacting dot. Moreover, for a completely isolated dot, the Green's function, Eq.~(\ref{GF_down}), assumes the same expression as for the Hubbard I approximation \cite{HewsonEOM}.

In the next step one has to establish a relation between Coulomb repulsion $U$ inside the dot and the induced superconducting gap $\Delta$, which we take as a unit in our considerations. By inspecting the experimental data for quantum dots embedded in the superconducting environment: InAs quantum dots \cite{Kanai,Lee2,Deng3} and InSb quantum dots \cite{Deng,Szombati}, we found  that in each case $U\gg\Delta$. This enables as to assume $U\rightarrow\infty$, which sets $A_{H\sigma}=1-\langle n_{\bar{\sigma}}\rangle$.

One notes that the Green's function spin $\sigma$ component depends on the QD occupancy $\langle n_{\bar{\sigma}}\rangle$, which has to be calculated selfconsistently. In equilibrium and finite temperature, it is performed through integration of the spectral density of the dot:
\begin{eqnarray}
\label{rho_sigma}
\rho_{\sigma}(\omega,\langle n_{\bar{\sigma}}\rangle)=-(1/\pi)\Im G_{\sigma}(\omega,\langle n_{\bar{\sigma}}\rangle),\\
\langle n_{\bar{\sigma}}\rangle=\int_{-\infty}^{+\infty}d\omega f(\omega)\rho_{\sigma}(\omega,\langle n_{\sigma}\rangle),
\end{eqnarray}
where $f(\omega)$ is Fermi distribution function.

\section{Results and discussion}

\subsection{Spin polarization of the Majorana bound states}

Let us introduce the spin polarization ascribed to $\gamma$ operator as:
\begin{equation}
\label{spin_pol}
P_{\gamma}=\frac{\langle\uparrow|n_{\gamma}|\uparrow\rangle-\langle\downarrow|n_{\gamma}|\downarrow\rangle}
{\langle\uparrow|n_{\gamma}|\uparrow\rangle+\langle\downarrow|n_{\gamma}|\downarrow\rangle},
\end{equation}
where $n_{\gamma}=\gamma^{\dagger}\gamma$ is a generalized occupancy operator and $|\sigma\rangle$ ($\sigma=\uparrow, \downarrow$) represent appropriate states with maximal spin up or down.

For the dot, taking into account Eq.~(\ref{d_composite}) for the generalized QD occupancy operator, we calculate dot's spin polarization within dot's states of spin-up and spin-down, and obtain $P_{dot}=1-2\cos^{2}(\Theta/4)$.

Similarly we proceed with the polarization of the Majorana states adjacent to the dot. Utilizing Eqs.(\ref{Majorana_s1}) and (\ref{Majorana_s2}),
the operators $n^{L}_{B,N,\sigma}=\gamma^{L}_{B,N,\sigma}\gamma^{L}_{B,N,\sigma}$ and $n^{R}_{A,1,\sigma}=\gamma^{R}_{A,1,\sigma}\gamma^{R}_{A,1,\sigma}$ are constructed. For the left Majorana spin components one obtains $\langle\uparrow |n^{L}_{B,N,\uparrow}|\uparrow\rangle=\sin^{2}(\Theta/4)\equiv\tau^{2}_{\uparrow}$ and $\langle\downarrow |n^{L}_{B,N,\downarrow}|\downarrow\rangle=\cos^{2}(\Theta/4)\equiv\tau^{2}_{\downarrow}$. Similarly we proceed with the right adjacent Majorana state $\gamma^{R}_{A,1}$. This yields spin polarizations $P_{\gamma^{L}_{B,N}}=P_{\gamma^{R}_{A,1}}=P_{dot}$.

Thus, the spin polarization of the Majoranas adjacent to the dot is determined by the dot's spin polarization, which in turn is determined by the resultant magnetic field inside the dot. This characteristic is supported by the picture emerging from several recent experiments; namely the sign of the Majorana spin polarization is determined by an external magnetic field or a  molecular field.

The former case is realized in topological wires \cite{Mourik}, where the direction of the spin of the active sub-band is determined by the external magnetic initiating topological state. It is similar in the case of a Majorana state located at the vortex center on the surface of a topological insulator in contact with an s-type superconductor \cite{Sun}; the spin polarization of the Majorana state is determined by the external field creating the vortex. As the authors demonstrate, it can be reversed by the change of the direction of magnetic field. Recently it has also been demonstrated theoretically, that the Majorana spin polarization in the vortex is totally parallel to the external magnetic field \cite{ChuangLi}.

The latter case is realized in an iron chain deposited on the Pb superconductor \cite{Jeon}, where the Majorana spin polarization is determined by the ferromagnetism of the chain and the specific spin sub-band crossing Fermi energy. It can also be reversed by reversing of the magnetization of STM tip or selecting a chain with different ferromagnetic orientation.

The uniqueness of the Majorana state polarization stems from the fact that it resides exactly on the Fermi level; this allows it to be distinguished from trivial Shiba states, which have resonances in both particle and hole domains with opposite polarizations, and when shifted towards Fermi energy, loose their spin polarization \cite{Cornils,Jeon}.

It is instructive to check whether the QD spin polarization can affect the utmost MBS, when hybridized with Majoranas in the vicinity of the dot, see Eq.~(\ref{H_hyb}). Taking into account Eq.~(\ref{f_alpha}) and the expressions for $\gamma^{L}_{B,N,\sigma}$ and $\gamma^{R}_{A,1,\sigma}$, Eqs.~(\ref{Majorana_s1}) and (\ref{Majorana_s2}), the spin-dependent components of the Majorana states lying at the farther ends of the wires  can be obtained as functions of the $\Theta$ angle:
\begin{eqnarray}
\label{MajoranaLA1}
\gamma^{L}_{A,1,\sigma}=2 f_{L\sigma}+\tau_{\sigma}(d^{\dagger}_{\sigma}-d_{\sigma}),\\
\label{MajoranaRBN}
\gamma^{R}_{B,N,\sigma}=-i[2 f_{R\sigma}+\tau_{\sigma}(d^{\dagger}_{\sigma}+d_{\sigma})],
\end{eqnarray}
where $\sigma=\downarrow$, $\uparrow$. By construction of the operators $n^{L}_{A,1,\sigma}$ and  $n^{R}_{B,N,\sigma}$ the spin polarizations of the MBS farther from the dot can be calculated, in the manner similar as for the MBS closer to the dot. The influence of the QD spin polarization on the farther MBS is now mediated by extended fermionic states $f_{\alpha\sigma}$. We obtain for the left and the right Majorana states  $\langle \sigma_{L}|n^{L}_{A,1,\sigma}|\sigma_{L}\rangle=4\langle \sigma_{L}|f^{\dagger}_{L\sigma}f_{L\sigma}|\sigma_{L}\rangle+\tau^{2}_{\sigma}$ and  $\langle \sigma_{R}|n^{R}_{B,N,\sigma}|\sigma_{R}\rangle=4\langle \sigma_{R}|f^{\dagger}_{R\sigma}f_{R\sigma}|\sigma_{R}\rangle+\tau^{2}_{\sigma}$, where $|\sigma_{\alpha}\rangle$ are the states of extended fermions $f_{\alpha}$ with spin $\sigma=\downarrow$ or $\uparrow$. The resultant spin polarization $P_{\gamma^{L}_{A,1}}=P_{\gamma^{R}_{B,N}}=\eta P_{dot}$ is reduced by the factor $\eta$ as compared to the dot polarization, and shows that the spin polarization of the distant Majorana states can also be controlled by the resultant magnetic field in the dot. The coefficient $\eta=1/9$  obtained in the present approach can be regarded as an rough estimation, which should depend on the distance between MBS within microscopic modelling.

This non-local control of the MBS spin polarization can be utilized for manipulation of the polarization of the current through the farther MBS.

\subsection{Density of states of the dot and transverse zero-bias conductance}

Utilizing  Eqs.~(\ref{GF_down}) and (\ref{rho_sigma}), the spin-$\sigma$ component of the transverse current  through the dot can be calculated within Meir-Wingreen \cite{meir} approach:
\begin{equation}
J_{\sigma}=\frac{e}{h}\frac{\pi\Gamma_{p\sigma}}{2}\int d\epsilon \left[f(\epsilon-\frac{e V}{2})-f(\epsilon+\frac{e V}{2})\right]\rho_{\sigma}(\epsilon).
\end{equation}

The current flows from the tunnel probe to the grounded superconductor. We assumed that the dot is weakly coupled to the normal probe, such that the entire voltage gradient is in the junction between the dot and the probe. For $V\rightarrow 0$ we obtain zero-bias conductance (ZBC):
\begin{equation}
\label{ZBC}
\mathcal{G}_{\sigma}=\frac{e^2}{h}\frac{\pi\Gamma_{p\sigma}}{2}\int d\epsilon \left (-\frac{\partial f(\epsilon)}{\partial\epsilon}\right )
\rho_{\sigma}(\epsilon).
\end{equation}

Our main interest concentrates on the sub-gap MBS behavior on Fermi level, $\epsilon_{F}=0$, so it is instructive to analyze the dot's spectral density features when $\omega\rightarrow \epsilon_{F}$. It will also immediately show the behavior of zero bias conductance in $T=0$, which directly probes the density of states of the dot at Fermi energy. Below  various cases with respect to MBS hybridizations in the wires are considered.

Firstly, let us consider the case of $\epsilon_{L}=0,\epsilon_{R}=0$, when the Majorana states adjacent to the dot  are not hybridized with their counter-partners at the opposite ends of the wires. Instead, they form the bound state $\epsilon_{d}$ inside the dot.

\begin{figure*} [ht]
\epsfxsize=0.65\textwidth
\epsfbox{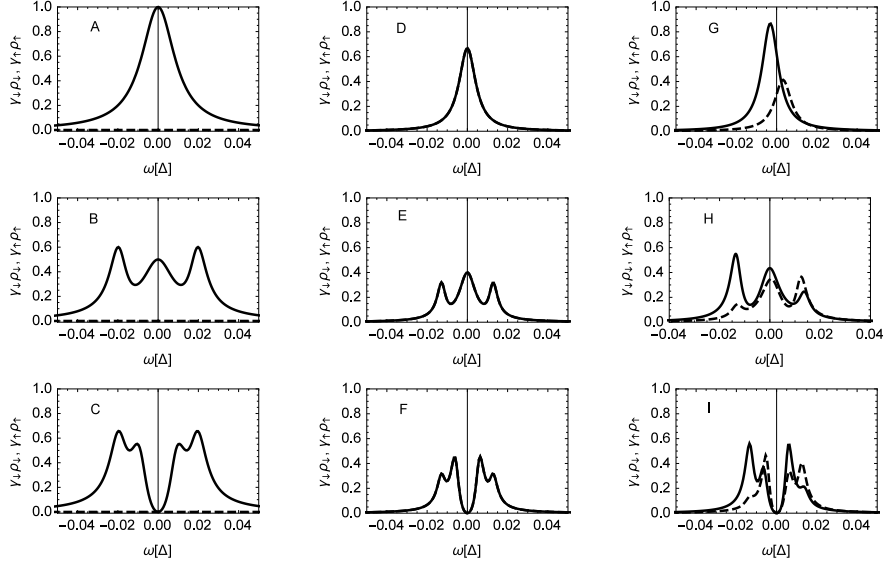}
\caption{\label{spec_densities} Spectral densities of the dot in spin-down (solid curves) and spin-up (dashed curves) sectors, multiplied by $\gamma_{\downarrow}=\Gamma_{p\downarrow}/2$ and $\gamma_{\uparrow}=\Gamma_{p\uparrow}/2$, respectively. The left column (A, B, C) is for $\Theta=0$, the middle column (D, E, F) is for $\Theta=\pi$ and the right column (G, H, I) is for $\Theta=0.99\pi$. The upper row (A, D, G) is for $\epsilon_{L}=\epsilon_{R}=0$, the middle row (B, E, H) is for $\epsilon_{L}=0.02$ and $\epsilon_{R}=0$ and the lower row (C, F, I) is for $\epsilon_{L}=0.02$ and $\epsilon_{R}=0.01$. The gate voltage is adjusted to obtain the resonance $\epsilon_{\downarrow}=\epsilon_{F}$ for $\Theta=0$ ($V_{g}=-2$) and for $\Theta=\pi$ ($V_{g}=0$) to yield $\epsilon_{\downarrow}=\epsilon_{\uparrow}=\epsilon_{F}$. For $\Theta=0.99\pi$ also $V_{g}=0$. Other input parameters are as follows: $\Gamma_{p}=0.02$, $B=2$, $t_{L}=t_{R}=0.1$ and $\Delta\phi=\pi$. }
\end{figure*}

For $\Theta=0$ only $\epsilon_{\downarrow}$ sublevel is active within the gap, whereas $\epsilon_{\uparrow}$ is pushed above Fermi level by the large Zeeman splitting. For the renormalized level $\tilde{\epsilon}_{\downarrow}$ in resonance with Fermi level, the limit $\omega\rightarrow \epsilon_{F}$ yields $\rho_{\downarrow}(\epsilon_{F})=(1/\pi\Gamma_{p\downarrow})(1-\langle n_{\uparrow}\rangle)$. It further yields the ZBC in $T=0$: $\mathcal{G}_{\downarrow}=\frac{e^2}{h}(1-\langle n_{\uparrow}\rangle)$, which results in the unitary limit of conductance because of the non-occupied upper sublevel $\langle n_{\uparrow}\rangle=0$. This feature is depicted in Figs.~\ref{spec_densities}A and \ref{ZBCfiniteT}A.

Consider now the case of $\Theta=\pi$, when the magnetic fields of the left and the right wires  cancel each other and the dot level retains its spin degeneracy. Taking into account Eqs.~(\ref{GF_down}) (\ref{rho_sigma}) and (\ref{ZBC}) for the resonant case $\tilde{\epsilon}_{\downarrow}=\tilde{\epsilon}_{\uparrow}=\epsilon_{F}$, we obtain the ZBC in $T=0$ for each spin direction:
\begin{equation}
\mathcal{G}_{\sigma}=\frac{e^2}{h}\frac{(1-\langle n_{\bar{\sigma}} \rangle)(\Gamma_{p\sigma}/2)^{2}}{(1/4)[(t_{L}+t_{R})\cos(\frac{\Delta\phi}{2})]^{2}+(\Gamma_{p\sigma}/2)^{2}}.
\end{equation}

Additionally, for $\Theta=\pi$: $\langle n_{\downarrow}\rangle=\langle n_{\uparrow}\rangle\equiv\langle n\rangle$ and $\Gamma_{p\downarrow}=\Gamma_{p\uparrow}$. Thus, the ZBC reaches its maximum with $4\pi$-period in $\Delta\phi$. Moreover, at its maximum, the occupancy in $T=0$,  for the resonant case, can easily be obtained analytically, yielding exactly $\langle n\rangle=1/3$. It implies the value of the ZBC $\mathcal{G}_{\sigma}=(2/3)\frac{e^2}{h}$, see  Figs.~\ref{spec_densities}D and \ref{ZBCfiniteT}D.  This result is a direct consequence of Coulomb interactions inside the dot, which is manifested by diminishing of the spectral weight $\sim(1-\langle n_{\bar{\sigma}}\rangle)$ of the QD spin sublevel resonance, as compared to the unitary weight in the non-interacting case. Taking into account Eq.~(\ref{composite_GF}), the total conductance at the resonance is $\mathcal{G}=\frac{e^2}{h}(1/2)(\mathcal{G}_{\downarrow}+\mathcal{G}_{\uparrow})=(2/3)\frac{e^2}{h}$.

The feature of $\langle n\rangle=1/3$ can also be observed for the first Hubbard peak, when in resonance with Fermi energy \cite{Stefanski_PRB}. We will show below, how this peculiar behavior transfers into the Majorana peak height. For compounds hosting mixed-valent Cerium ions, similar tendency of the lowering level occupancy and diminishing the height of the spectral density peaks can be observed, when density of states is calculated within the local spin density approximation with Coulomb interactions (LSDA+U), as compared to LSDA without interactions, see for instance \cite{Kaczorowski}.

Consider now the case, when the hybridization between the end-Majorana states in the wire $\alpha$ is $\epsilon_{\alpha}\neq 0$, whereas in the other wire is $\epsilon_{\alpha'}=0$. In such an arrangement the unpaired Majorana state at the end of the $\alpha'$-wire, adjacent to the dot, produces a characteristic peak in the spectral density of the dot at $\omega=\epsilon_{F}$. The $\sigma$-part of ZBC  in $T=0$ assumes the form:
\begin{equation}
\label{ZBCdoT0}
\mathcal{G}_{\sigma}=\frac{e^{2}}{h} \frac{(1/4)(1-\langle n_{\bar{\sigma}}\rangle)(2-\langle n_{\bar{\sigma}}\rangle)\Gamma^{2}_{p\sigma}}
{(1/4)(2-\langle n_{\bar{\sigma}}\rangle)^{2}\Gamma^{2}_{p\sigma}+\langle n_{\bar{\sigma}}\rangle^{2}\tilde{\epsilon}_{\sigma}}.
\end{equation}

For $\Theta=0$, $\langle n_{\uparrow}\rangle=0$ and $\mathcal{G}_{\downarrow}=\frac{e^{2}}{2h}$ reproduces the non-interacting limit. This result is independent of the position of $\epsilon_{\downarrow}$.

For $\Theta=\pi$ and the resonant case, when $\tilde{\epsilon}_{\downarrow}=\tilde{\epsilon}_{\uparrow}=\epsilon_{F}$, Eq.~(\ref{ZBCdoT0}) yields:
\begin{equation}
\label{G_limit}
\mathcal{G}_{\sigma}=\frac{e^{2}}{h}\frac{1-\langle n_{\bar{\sigma}}\rangle}{2-\langle n_{\bar{\sigma}}\rangle.}
\end{equation}

In the present case the occupancy of the dot cannot be as easily calculated analytically as previously. However, from the inspection of the selfconsistently found occupancies vs. gate voltage it follows  that  the cases of $\Theta=\pi$ for $\epsilon_{L}, \epsilon_{R}=0$ and $\Theta=\pi$ for  $\epsilon_{\alpha}\neq 0, \epsilon_{\alpha'}=0$ are virtually indistinguishable. Thus, the formation of the extended fermionic states in the wires does not alter considerably the dot occupancy. This allows us to assume $\langle n_{\bar{\sigma}}\rangle\cong 1/3$, as previously, which yields the value of the Majorana peak height in zero bias conductance: $\mathcal{G}_{\sigma}\cong (2/5)\frac{e^2}{h}$ per spin. This is the result of electron interactions inside the dot. Then the total conductance at the peak amounts to $\mathcal{G}\cong (2/5)\frac{e^2}{h}$, which is diminished as compared to the half of conductance for the non-interacting dot and only the spin-$\downarrow$ sector active.

Generally the height of the Majorana peak in ZBC of $\sigma$-sector is scaled approximately by $\sim(1-\langle n_{\bar{\sigma}}\rangle)$, and its maximal value is half conductance quantum  for $\Theta=0$.

The last case remaining to be considered is for the fully paired Majoranas in each wire, which yields in $T=0$ the ZBC $\mathcal{G}_{\sigma}=0$ independently of the $\epsilon_{\sigma}$ position and the $\Theta$ angle.

In Fig.~\ref{spec_densities} spectral densities of the dot, multiplied by $\gamma_{\sigma}=\Gamma_{p\sigma}/2$,  are displayed for various resultant effective magnetic fields inside the dot and different hybridizations of the Majorana states in the wires.

The left column corresponds to $\Theta=0$ and the maximal Zeeman splitting of the dot level. Fig. \ref{spec_densities}A, \ref{spec_densities}B and \ref{spec_densities}C correspond to the cases of $\epsilon_{L}=\epsilon_{R}=0$, $\epsilon_{L}\neq 0$, $\epsilon_{R}=0$ and $\epsilon_{L}\neq 0$, $\epsilon_{R}\neq 0$, respectively. This sequence is also retained for the middle column- Figs.~\ref{spec_densities}D, E, F, which corresponds to for $\Theta=\pi$, and the right column- Figs.~\ref{spec_densities}G, H, I for $\Theta=99\pi$.   For  $\epsilon_{L}=\epsilon_{R}=0$, the Majoranas adjacent to the dot  in the left and the right wire hybridize inside the dot, forming a bound state. For $\Theta=0$ the $\epsilon_{\uparrow}$ sublevel is pushed above Fermi energy, whereas $\epsilon_{\downarrow}$, when in resonance with Fermi energy, produces a unitary peak in spin-down spectral density. As shown above, the unitary limit is reached despite of electron interactions as the $\epsilon_{\uparrow}$ is unoccupied, yielding $\langle n_{\uparrow}\rangle=0$. In Fig.~\ref{spec_densities}B, for $\epsilon_{L}\neq 0$, $\epsilon_{R}=0$, the MBS in the left wire are paired, whereas the unpaired Majorana state at the end of the right wire, close to the dot, produces a characteristic resonance at $\omega=0$. It reaches half-unity value, which transfers into half conductance quantum in ZBC in $T=0$ (see also Fig.~\ref{ZBCfiniteT}B for finite temperature). The height of this peak is independent of the $\epsilon_{\downarrow}$ position.  For $\epsilon_{L}\neq 0$ and $\epsilon_{R}\neq 0$ in Fig.~\ref{spec_densities}C, the Majoranas of the left and the right wires are paired, forming extended fermionic states, seen as satellite resonances at $\omega\sim\mp\epsilon_{L}$ and  $\omega\sim\mp\epsilon_{R}$. In turn, the Majorana resonance disappears completely and $\rho_{\downarrow}(\omega=\epsilon_{F})=0$, independently of $\epsilon_{\downarrow}$ position. In general, the behavior of the spin-down spectral density for $\Theta=0$ resembles the case of the effectively non-interacting dot with spin-down  sublevel active \cite{PStefanskiJPCM16}.

In Figs.~\ref{spec_densities}D, E, F  spectral densities for $\Theta=\pi$ are shown, when the dot level is spin-degenerate. In Fig.~\ref{spec_densities}D,  for  $\epsilon_{L}=\epsilon_{R}=0$, a resonance is produced at Fermi level when $\tilde{\epsilon}_{\downarrow}=\tilde{\epsilon}_{\uparrow}=\epsilon_{F}$. As derived above, the height of the resonance is diminished by electron interactions as compared to the quasi-noninteracting case, and in this arrangement it is exactly $2/3$. Fig.~\ref{spec_densities}E corresponds to  $\epsilon_{L}\neq 0$ and $\epsilon_{R}=0$ with the spectral densities displaying Majorana peak at $\omega=0$. The height of this peak in this arrangement is $\sim 2/5$, as discussed above. The two satellite peaks originating from the Majorana-hybridized extended fermion in the left wire are shifted towards $\omega=0$, as compared to the quasi-noninteracting dot (Fig.~\ref{spec_densities}B), see also the discussion below.

The discussed features of the spectral density of the dot can be experimentally revealed by examination of the transverse differential conductance trough the dot. In the following we focus on the presentation of the transverse zero-bias conductance.
\begin{figure} [ht]
\epsfxsize=0.3\textwidth
\epsfbox{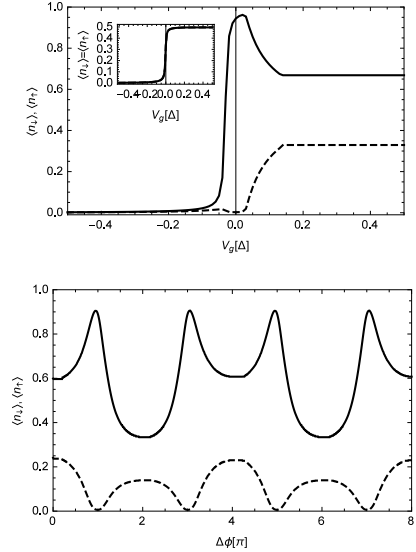}
\caption{\label{particle_nr} Upper panel: selfconsistently calculated QD occupancies vs. gate voltage  for spin-down (solid curves) and spin-up (dashed lines) sectors, evaluated in $T=0.01$ for $\Theta=0.99 \pi$ and $\Delta\phi=\pi$. The inset corresponds to the spin degeneracy, $\Theta=\pi$. Lower panel: QD occupancies vs. superconducting phase difference calculated for $\Theta=0.99 \pi$ and $V_{g}=0$. Other input parameters are as follows: $\Gamma_{p}=0.02$, $B=2$, $t_{L}=t_{R}=0.1$, $\epsilon_{L}=\epsilon_{R}=0$. }
\end{figure}

In Fig.~\ref{particle_nr}  the representative dependencies of the dot occupancies vs. gate voltage (upper panel) and vs. phase difference for the set gate voltage (lower panel) are shown. The inset displays the occupancies calculated  for spin degeneracy, $\Theta=\pi$. For a given $\Theta$ value the dependencies on the gate voltage are similar, independently of the sequence of hybridizations $\epsilon_{L}$ and $\epsilon_{R}$. For the maximal Zeeman splitting, $\Theta=0$ (not shown), the $\langle n_{\downarrow}\rangle$ dependence is similar as in the inset, but for large gate voltages the occupancy approaches unity, whereas $\langle n_{\uparrow}\rangle\sim 0$, independently of the  gate voltage. The characteristic $4\pi$ periodicity of the occupancies vs. $\Delta\phi$, shown in the lower panel, is a consequence of the MBS existence in the junction, and specifically of the $\Delta\phi$ dependencies of the renormalized $\tilde{\epsilon}_{\downarrow}$ and $\tilde{\epsilon}_{\uparrow}$ sub-levels. In the present configuration ($V_{g}=0$), for odd values of $\Delta\phi$, the sub-levels realize the symmetrical alignment  $\tilde{\epsilon}_{\downarrow}<\epsilon_{F}< \tilde{\epsilon}_{\uparrow}$. For even values of   $\Delta\phi$ both the levels are pushed above and below Fermi level interchangeably, starting from below $\epsilon_{F}$ for $\Delta\phi=0$. The maxima in $\langle n_{\downarrow}\rangle$ dependence are accompanied by the minima of $\langle n_{\uparrow}\rangle$ and vice versa, which is a result of electron repulsion inside the dot.

\begin{figure*} [ht]
\epsfxsize=0.65\textwidth
\epsfbox{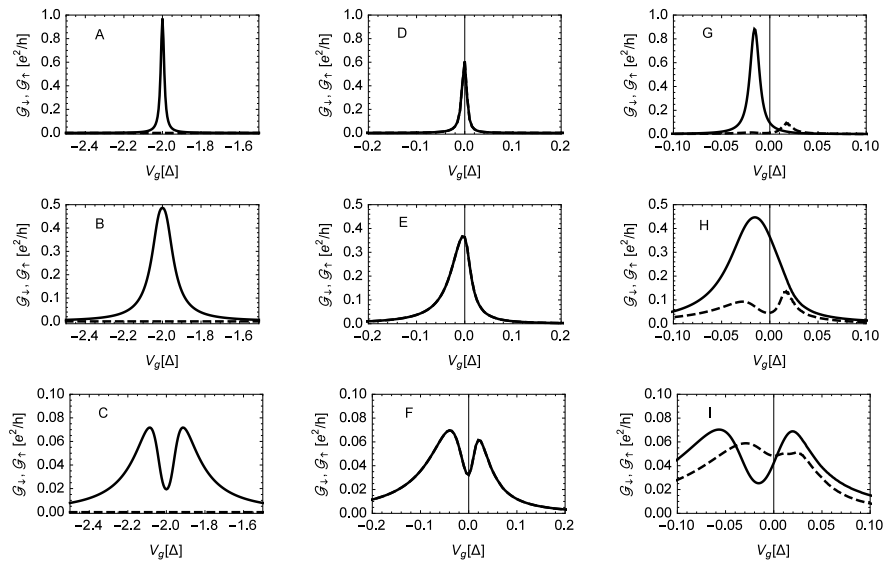}
\caption{\label{ZBCfiniteT} Transverse zero-bias conductances through the dot in the spin-down (solid curves) and the spin-up (dashed curves) sectors, calculated in $T=0.01$. The left column (A, B C) corresponds to $\Theta=0$, the middle column (D, E, F) corresponds to $\Theta=\pi$ and the right column (G, H, I) corresponds to $\Theta=0.99\pi$. The upper row (A, D, G) is for $\epsilon_{L}=\epsilon_{R}=0$, the middle row (B, E, H) corresponds to $\epsilon_{L}=0.02$ and $\epsilon_{R}=0$ and the lower row (C, F, I) corresponds to $\epsilon_{L}=0.02$ and $\epsilon_{R}=0.01$. Other input parameters are the same as in Fig.~\ref{spec_densities}.}
\end{figure*}

In Fig.~\ref{ZBCfiniteT}  transverse zero-bias conductances through the dot, calculated in temperature $T=0.01$, are shown. The sequence of panels is the same as for Fig.~\ref{spec_densities}. The left column for $\Theta=0$, which corresponds to the largest Zeeman splitting and the quasi-noninteracting case, is presented for comparison purposes. One notes that $\epsilon_{\downarrow}$ is situated deep below Fermi energy and the application of the large negative voltage, $V_{g}=-2$, tunes it into resonance with Fermi energy. For $\Theta=\pi$ (the middle column) and $\Theta=0.99 \pi$ (the right column) electron interactions are profoundly manifested in conductance. Firstly, for $\Theta=\pi$ and $\epsilon_{L}=\epsilon_{R}=0$ (Fig.~\ref{ZBCfiniteT}D) the heights of the conductance peaks are diminished, as compared to the $\Theta=0$ case, as discussed above. Secondly,  the conductance peak originating from the unpaired Majorana state  exhibits a characteristic asymmetry, (Fig.~\ref{ZBCfiniteT}E). It is caused by the change of the spectral weight $\sim(1-\langle n_{\bar{\sigma}}\rangle)$ of the QD spectral density in the $\sigma$-sector. For negative (positive) gate voltages, where the occupancy is very small (it reaches $1/2$) (see the inset in the upper panel of Fig.~\ref{particle_nr}), the spectral weight is larger (smaller), which produces the asymmetry of the conductance peak. Also the asymmetry, caused by the same effect, is seen in (Fig.~\ref{ZBCfiniteT}F), for $\epsilon_{L}\neq 0, \epsilon_{R}\neq 0$, when all the Majoranas are paired. This parity effect, induced by electron interactions in the dot, is gradually diminished with the increase of temperature,  as a result of smoothing of the occupancy dependencies vs. gate voltage (not shown). It is worth to emphasizing that the Majorana peak itself, emerging in the spectral density of the dot, remains symmetric with the change of gate voltage (not shown).

The asymmetry of the conductance peak has also been predicted \cite{vanHeck} in the transport through a Coulomb blockaded proximitized wire in the limit of sequential tunneling, whereas the peak due to the conductance via the Majorana states remained symmetric.

In the case of a quasi-noninteracting limit of the present model, depicted in the left column, the peaks in conductance are symmetric, since $\langle n_{\uparrow}\rangle\sim 0$ independently of the gate voltage, and the renormalization of the spectral weight by $\sim(1-\langle n_{\uparrow}\rangle)$ has no effect.

Electron interactions are  also strikingly manifested for a finite Zeeman splitting, depicted in the right column of Fig.~\ref{ZBCfiniteT}. The peculiar behavior of the occupancies at $V_{g}\sim 0$ in this case (see the upper panel of  Fig.~\ref{particle_nr}), namely the large value of $\langle n_{\downarrow}\rangle$  and the small value of $\langle n_{\uparrow}\rangle$, pushes the spin-up conductance peak, caused by the Majorana state, toward positive gate voltages, whereas the peak in spin-down sector is shifted toward negative gate voltages (shown in Fig.~\ref{ZBCfiniteT}H). The large asymmetry and shifting of the peaks in (Fig.~\ref{ZBCfiniteT}I) for $\epsilon_{L}\neq0, \epsilon_{R}\neq 0$ are caused by the same parity effect.
We believe that presented  diminishing of the Majorana peak in the conductance, caused by electron interactions, can be spotted experimentally due to unprecedented precision of the recent transport spectroscopy measurements \cite{Zhang}.

\subsection{Renormalization of Majorana splitting}
From the inspection of the expressions for the spectral density of the dot in the spin-up and the spin-down sectors, following from Eq.~(\ref{GF_down}), and their comparison to the expressions for the non-inteacting case, it can be noticed, that the hybridization of $\epsilon_{\alpha}$ in the wire $\alpha$ is altered by the presence of electron interactions and the finite magnetic field inside the dot. The hybridization in the $\alpha$-wire now has  spin components:  $\tilde{\epsilon}_{\alpha}=\epsilon_{\alpha\downarrow}+\epsilon_{\alpha\uparrow}$  resulting from the QD spin sublevels  distinguishable by Zeeman splitting, each of them converted from the pair of the MBS spin components closer to the dot. However, as a consequence of spin mixing, they should be rather regarded as a non-separable sum.
Having this in mind, we can analyse the behavior of the spin components. Namely,  in the $\alpha$-wire the spin-down component  is renormalized as $\epsilon_{\alpha\downarrow}=\epsilon_{\alpha}\cos^{2}(\Theta/4)(1-\langle n_{\uparrow}\rangle)$, whereas the spin-up component is $\epsilon_{\alpha\uparrow}=\epsilon_{\alpha}\sin^{2}(\Theta/4)(1-\langle n_{\downarrow}\rangle)$. For small finite Zeeman splitting, as shown in Figs.~\ref{spec_densities}G, H, I for $\Theta=0.99\pi$, the following QD sub-levels arrangement occurs $\epsilon_{\downarrow}<\epsilon_{F}<\epsilon_{\uparrow}$ and $\langle n_{\uparrow}\rangle<\langle n_{\downarrow}\rangle$. This implies that the renormalized hybridizations are in relation  $\tilde{\epsilon}_{\alpha\downarrow}>\tilde{\epsilon}_{\alpha\uparrow}$, which is reflected in Fig.~\ref{spec_densities}H and in Fig.~\ref{spec_densities}I, where the satellite peaks of the extended fermionic states in the spin-down sector are more apart from each other on the energy scale than  the corresponding peaks in the spin-up sector.

Let us examine some simple limits of $\tilde{\epsilon}_{\alpha}$. i) For $\Theta=0$ and a quasi-non-interacting dot it is clear that $\tilde{\epsilon}_{\alpha}=\epsilon_{\alpha\downarrow}$, where the spin index can be omitted as the only spin index in this case. ii)  For $\Theta=\pi$ (QD spin degeneracy) $\tilde{\epsilon}_{\alpha}=\epsilon_{\alpha}(1-\langle n\rangle)$, where $\langle n\rangle\equiv\langle n_{\downarrow}\rangle=\langle n_{\uparrow}\rangle$. In this case, the MBS hybridization reveals its dependence on the parity of the junction. For large negative gate voltages, when the dot's level is unoccupied, the MBS hybridization retains its value as for non-interacting case. However, for large positive gate voltages, when the dot's level is occupied and $\langle n\rangle=1/2$, the hybridization is diminished to the value $\epsilon_{\alpha}/2$. This effect can be ascribed to the competition between hybridization of the end-MBS in each wire and hybridization of the MBS states through the dot, the latter being enhanced when the converted dot level is occupied. It should be emphasized that this effect is operative only in presence of  interactions in the  dot. iii) For a small Zeeman splitting, such that $\cos^{2}(\Theta/4)\sim\sin^{2}(\Theta/4)\sim 1/2$, utilizing the condition for total occupancy, $\langle n_{\downarrow}\rangle+\langle n_{\uparrow}\rangle=1$, yields $\tilde{\epsilon}_{\alpha}\sim\epsilon_{\alpha}/2$. iv) Finally, for an arbitrary Zeeman splitting and both unoccupied QD spin-sublevels, the hybridizations have values as in the non-interacting case.

It has been shown within more realistic models \cite{Das Sarma2,klin2} that Majorana splitting exhibits an oscillatory behavior with respect to the magnetic field (the chemical potential) for the fixed chemical potential (magnetic field) in the wire, which can be a "smoking gun" of the Majorana existence. Both the parameters induce an oscillatory behavior of the charge density in the wire, which affects the hybridizations of the Majorana wave functions. This feature may be obscured by Coulomb interactions, which selfconsistently compensate the changes in the charge density induced by the change of the magnetic field (or the chemical potential).

In the present model, the change of the dot density (particle number) exhibits $4\pi$ periodicity in each spin sector, characteristic for the MBS  created Josephson junction, depicted in the lower panel of Fig.~\ref{particle_nr}. However, due to the dependence on the sum of spin occupancies, this oscillation will be hardly visible in the splitting of the Majorana resonance. Generally it follows that the Majorana splitting is diminished by electron interactions in the dot.

\section{Diagonalization of the Hamiltonian in Nambu space}

In order to analyse the energy structure of the junction and the phase-biased Josephson current, the Hamiltonian is rewritten in the generalized Nambu basis and diagonalized numerically. In this step we neglect the influence of the tunnelling electrode, which can be decoupled from the dot when the Josephson current is measured. We also restrict ourselves to the case of $U\rightarrow\infty$, which allows the dot level to be occupied by at most one electron. It implies that the term $\sim U$ in Eq.~(\ref{Hdot}), which is not quadratic, drops down.

Thus, let us rewrite the Hamiltonian described by Eqs.~(\ref{Hdot}), (\ref{H_tun_simp_fer}) and (\ref{H_ov_fer})  within the generalized Nambu basis $\Psi=(d_{\downarrow},d_{\uparrow},f_{L\downarrow},f_{L\uparrow},f_{R\downarrow},f_{R\uparrow},d^{\dagger}_{\uparrow},d^{\dagger}_{\downarrow},f_{L\uparrow}^{\dagger},
f_{L\downarrow}^{\dagger},f_{R\uparrow}^{\dagger},f_{R\downarrow}^{\dagger})$. It can be written as:
\begin{equation}
H_{s}=\frac{1}{2}\Psi^{\dagger}\tilde{H_{s}}\Psi+\frac{1}{2}(t_{L}+t_{R})\cos\left(\frac{\Delta\phi}{2}\right)
\end{equation}
with
\begin{widetext}
\begin{equation}
\label{H_matrix}
\tilde{H_{s}}=\left(
\begin{array}{cccccccccccc}
\tilde{\epsilon}_{\downarrow} & -t\alpha\beta & -\alpha\frac{\epsilon_{L}}{2} & 0 & -\alpha\frac{\epsilon_{R}}{2} & 0 & 0 & 0 & 0 & -\alpha\frac{\epsilon_{L}}{2} & 0 & \alpha\frac{\epsilon_{R}}{2}\\
-t\alpha\beta & \tilde{\epsilon}_{\uparrow} & 0 & -\beta\frac{\epsilon_{L}}{2} & 0 & -\beta\frac{\epsilon_{R}}{2} & 0 & 0 & -\beta\frac{\epsilon_{L}}{2} & 0 & \beta\frac{\epsilon_{R}}{2} & 0\\
-\alpha\frac{\epsilon_{L}}{2} & 0 & 0 & 0 & 0 & 0 & 0 & \alpha\frac{\epsilon_{L}}{2} & 0 & 0 & 0 & 0\\
0 & -\beta\frac{\epsilon_{L}}{2} & 0 & 0 & 0 & 0 & \beta\frac{\epsilon_{L}}{2} & 0 & 0 & 0 & 0 & 0\\
-\alpha\frac{\epsilon_{R}}{2} & 0 & 0 & 0 & 0 & 0 & 0 & -\alpha\frac{\epsilon_{R}}{2} & 0 & 0 & 0 & 0\\
0 & -\beta\frac{\epsilon_{R}}{2} & 0 & 0 & 0 & 0 & -\beta\frac{\epsilon_{R}}{2} & 0 & 0& 0 & 0 & 0\\
0 & 0 & 0 & \beta\frac{\epsilon_{L}}{2} & 0 & -\beta\frac{\epsilon_{R}}{2} &  -\tilde{\epsilon}_{\uparrow} &  t\alpha\beta & \beta\frac{\epsilon_{L}}{2} & 0 &
\beta\frac{\epsilon_{R}}{2} & 0\\
0 & 0  & \alpha\frac{\epsilon_{L}}{2} & 0 & -\alpha\frac{\epsilon_{R}}{2} & 0 & t\alpha\beta & -\tilde{\epsilon}_{\downarrow} & 0 & \alpha\frac{\epsilon_{L}}{2} & 0 & \alpha\frac{\epsilon_{R}}{2}\\
0 & -\beta\frac{\epsilon_{L}}{2} & 0 & 0 & 0 & 0 & \beta\frac{\epsilon_{L}}{2} & 0 & 0 & 0 & 0 & 0\\
-\alpha\frac{\epsilon_{L}}{2} & 0 & 0 & 0 & 0 & 0 & 0 & \alpha\frac{\epsilon_{L}}{2} & 0 & 0 & 0 & 0\\
0 & \beta\frac{\epsilon_{R}}{2} & 0 & 0 & 0 & 0 & \beta\frac{\epsilon_{R}}{2} & 0 & 0 & 0 & 0 & 0\\
\alpha\frac{\epsilon_{R}}{2} & 0 & 0 & 0 & 0 & 0 & 0 & \alpha\frac{\epsilon_{R}}{2} & 0 & 0 & 0 & 0\\
\end{array}
\right).
\end{equation}
\end{widetext}
The eigen-energy structure of the Hamiltonian reveals  Andreev bound states (ABS) with their quasiparticle component $E^{ABS}_{i+}$ and its quasihole counterpart $E^{ABS}_{i-}=-E^{ABS}_{i+}$. In our case we also encounter "parti-hole" levels positioned at Fermi energy, which possess the properties of the Majorana bound states (MBS).

It is also useful for further analysis to ascribe to the $i$-th ABS level the second quantized operator $\gamma_{i}=\Psi\phi_{i}$, where $\phi_{i}$ is the eigenvector associated with this ABS level.

\subsection{Contribution of the Majorana bound states to the spectral density of the dot, determined from the Hamiltonian's spectrum}
It is instructive to analyze the contributions to the density of states originating from different ABS levels, and in particular from the Majorana bound states. To do so we start from the Lehmann representation of the QD Green's function \cite{Negele, Bulla}, which in $T=0$ has the particle and hole parts:
\begin{eqnarray}
\label{rho+}
\rho^{+}_{\sigma}(\omega)=\sum_{i}|\langle\phi_{i}|d^{\dagger}_{\sigma}|g\rangle|^{2}\delta(\omega-(E_{i}^{ABS}-E_{g})),\\
\label{rho-}
\rho^{-}_{\sigma}(\omega)=\sum_{i}|\langle g|d^{\dagger}_{\sigma}|\phi_{i}\rangle|^{2}\delta(\omega+(E_{i}^{ABS}-E_{g})),
\end{eqnarray}
where the subscript $g$ denotes ground state ($E_{g}=0$), and the set of $|\phi_{i}\rangle$ and  $E_{i}^{ABS}$ ($i=1,..,12$) constitutes the eigen-spectrum of the Hamiltonian. In our discussion it is convenient to rewrite Eqs.~(\ref{rho+}) and (\ref{rho-}) in the terms of the second quantized operators $\gamma_{i}$ corresponding to  the $E_{i}^{ABS}$ levels. First Eqs.~(\ref{rho+}) and (\ref{rho-}) are written in a different form:
\begin{eqnarray}
\label{rho+}
\rho^{+}_{\sigma}(\omega)=\sum_{i}|\langle\phi_{i}|d^{\dagger}_{\sigma}d_{\sigma}|\phi_{i}\rangle|^{2}\delta(\omega-E_{i}^{ABS}),\\
\label{rho-}
\rho^{-}_{\sigma}(\omega)=\sum_{i}|\langle\phi_{i}|d_{\sigma}d^{\dagger}_{\sigma}|\phi_{i}\rangle|^{2}\delta(\omega+E_{i}^{ABS}),
\end{eqnarray}
and then the operators $d^{\dagger}_{\sigma}d_{\sigma}$ and $d_{\sigma}d^{\dagger}_{\sigma}$ are rewritten in the dot occupation number  basis, which yields the Hubbard operators \cite{Coleman}: $d^{\dagger}_{\sigma}d_{\sigma}=|1\rangle_{\sigma}\langle 1|_{\sigma}$ and $d_{\sigma}d_{\sigma}^{\dagger}=|0\rangle_{\sigma}\langle 0|_{\sigma}$. Finally, we switch from the eigenvectors $|\phi_{i}\rangle$ notation to the corresponding $\gamma_{i}$ operators notation:
\begin{eqnarray}
\label{rho+op}
\rho^{+}_{\sigma}(\omega)=\sum_{i}\gamma_{i}^{\dagger}|1\rangle_{\sigma}\langle 1|_{\sigma}\gamma_{i}\delta(\omega-E_{i}^{ABS}),\\
\label{rho-op}
\rho^{-}_{\sigma}(\omega)=\sum_{i}\gamma_{i}^{\dagger}|0\rangle_{\sigma}\langle 0|_{\sigma}\gamma_{i}\delta(\omega+E_{i}^{ABS}).
\end{eqnarray}
Taking into account the general view of the $\gamma_{i}$ operator corresponding to a given $i-$th ABS level:
\begin{widetext}
\begin{eqnarray}
\label{ABSoperator}
\nonumber
\gamma_{i}=\Psi\phi_{i}=a_{i,1}d_{\downarrow}+a_{i,2}d_{\uparrow}+a_{i,3}f_{L\downarrow}+a_{i,4}f_{L\uparrow}+a_{i,5}f_{R\downarrow}+a_{i,6}f_{R\uparrow}+\\
a_{i,7}d_{\uparrow}^{\dagger}+a_{i,8}d_{\downarrow}^{\dagger}+a_{i,9}f_{L\uparrow}^{\dagger}+a_{i,10}f^{\dagger}_{L\downarrow}+
a_{i,11}f_{R\uparrow}^{\dagger}+a_{i,12}f_{R\downarrow}^{\dagger},
\end{eqnarray}
\end{widetext}
we arrive at the following equations:
\begin{eqnarray}
\label{rhodown+}
\rho^{+}_{\downarrow}(\omega)=\sum_{i} |a_{i,8}|^{2}\delta(\omega-E_{i}^{ABS}),\\
\label{rhodown-}
\rho^{-}_{\downarrow}(\omega)=\sum_{i} |a_{i,1}|^{2}\delta(\omega+E_{i}^{ABS}),\\
\label{rhoup+}
\rho^{+}_{\uparrow}(\omega)=\sum_{i} |a_{i,7}|^{2}\delta(\omega-E_{i}^{ABS}),\\
\label{rhoup-}
\rho^{-}_{\uparrow}(\omega)=\sum_{i} |a_{i,2}|^{2}\delta(\omega+E_{i}^{ABS}).
\end{eqnarray}
The above derivation is a generalization of the analogous equations for the simple BCS Hamiltonian \cite{Matsui} concerning more complex quasiparticles.

We are mainly interested in the determination of the contributions to the QD spectral density originating from the unpaired Majorana states. Thus, let us analyze in detail the arrangement of $\epsilon_{L}\neq 0$ and $\epsilon_{R}=0$, when  the end-state Majoranas in the left wire form an extended fermionic state $f_{L}$, whereas the MBS adjacent to the dot in the right wire is left unpaired.

For $\Theta=0$, when only the spin-$\downarrow$ sector is active, there are two states residing at Fermi level:
\begin{eqnarray}
E_{1}^{ABS}=0, & \gamma_{1\downarrow}=\frac{i}{\sqrt{2}}(f^{\dagger}_{L\downarrow}-f_{L\downarrow}),\\
E_{2}^{ABS}=0, & \gamma_{2\downarrow}=\frac{1}{\sqrt{2+(\frac{2\tilde{\epsilon}_{\downarrow}}{\epsilon_{L}})^{2}}}(d^{\dagger}_{\downarrow}+d_{\downarrow}+
\frac{2\tilde{\epsilon}_{\downarrow}}{\epsilon_{L}}f_{L\downarrow}).
\end{eqnarray}
$\gamma_{1\downarrow}$ is the Majorana operator of the (paired) MBS, adjacent to the dot in the left wire, whereas  $\gamma_{2\downarrow}$ is the unpaired MBS with some admixture of the extended $f_{L}$ fermionic state. It has its contribution to the spin-down spectral density of the dot. According to Eqs. (\ref{rhodown+}) and (\ref{rhodown-}), we obtain $\rho_{\downarrow}(\epsilon_{f})=\rho^{+}_{\downarrow}(\epsilon_{f})+\rho^{-}_{\downarrow}(\epsilon_{f})=1/(\pi\Gamma_{\downarrow})$ when $\tilde{\epsilon}_{\downarrow}\rightarrow\epsilon_{F}$. It yields half of conductance quantum in the transverse zero-bias conductance: $\mathcal{G}_{\downarrow}=e^{2}/(2h)$ in $T=0$.

Consider now the case of an arbitrary angle $\Theta$. For $\epsilon_{R}=0$ the considered Nambu space has a reduced dimension due to the absence of the extended $f_{R}$ fermionic state. From diagonalization of the Hamiltonian we obtain the following states at Fermi energy:
\begin{eqnarray}
\label{Maj1}
E_{1}^{ABS}=0, & \gamma_{1\downarrow}=\frac{i}{\sqrt{2}}(f^{\dagger}_{L\downarrow}-f_{L\downarrow}),\\
E_{2}^{ABS}=0, & \gamma_{2\uparrow}=\frac{i}{\sqrt{2}}(f^{\dagger}_{L\uparrow}-f_{L\uparrow}),\\\label{Maj3}
E_{3}^{ABS}=0, & \gamma_{3\downarrow}=d_{\downarrow}+\frac{2(\epsilon_{\downarrow}-t\alpha^2)}{\alpha\epsilon_{L}}f_{L\downarrow}-\frac{2t\alpha}{\epsilon_{L}}f_{L\uparrow}
+d_{\downarrow}^{\dagger},\\\label{Maj4}
E_{4}^{ABS}=0, & \gamma_{4\uparrow}=d_{\uparrow}-\frac{2t\beta}{\epsilon_{L}}f_{L\downarrow}+\frac{2(\epsilon_{\uparrow}-t\beta^2)}{\beta\epsilon_{L}}f_{L\uparrow}
+d_{\uparrow}^{\dagger},
\end{eqnarray}
where the coefficients in the expressions for $\gamma_{3}$ and $\gamma_{4}$ operators are left unnormalized in order to save space.

The operators $\gamma_{1\downarrow}$ and $\gamma_{2\uparrow}$  are the spin components of the paired Majorana state $\gamma^{L}_{B,N}$ and have zero spin polarization. The operators $\gamma_{3\downarrow}$ and $\gamma_{4\uparrow}$ represent the spin components of the unpaired MBS, $\gamma^{R}_{A,1}$, and have some admixture of the extended $f_{L\sigma}$ fermionic states. It follows from the inspection of the coefficients in Eqs.~(\ref{Maj3}) and (\ref{Maj4}), that the admixtures of $f_{L\sigma}$ states vanish in both the spin sectors at the same time for the spin degeneracy: $\Theta=\pi$ and for $\tilde{\epsilon}_{\downarrow}=\tilde{\epsilon}_{\uparrow}=\epsilon_{F}$. For such an arrangement $\gamma_{3\downarrow}$ and $\gamma_{4\uparrow}$ become the Majorana bound states strictly localized at the dot.

Utilizing Eqs.~(\ref{rhodown+})-(\ref{rhoup-}), the contributions to the density of states of the dot, originating from $\gamma_{3\downarrow}$ and $\gamma_{4\uparrow}$ levels, are obtained. For $\Theta=\pi$, when  $\tilde{\epsilon}_{\downarrow}=\tilde{\epsilon}_{\uparrow}=\epsilon_{F}$ it yields $\rho_{3\downarrow/4\uparrow}(\epsilon_{F})=1/(\pi\Gamma_{\downarrow/\uparrow})$, which implies a half of conductance quantum $\mathcal{G}_{\downarrow/\uparrow}=e^{2}/(2h)$ in each spin sector. It can be confronted with the result for an interacting dot, Eq.~(\ref{G_limit}), which vields $\sim (2/5)e^{2}/(2h)$ per spin in this arrangement. Note that neglecting the interaction term in the diagonalized Hamiltonian yields the results for the non-interacting dot, thus the diminishing effect of electron correlations on the Majorana peak is not reflected here.

Let us introduce the quantities describing the degree of localization inside the dot of the MBS $\gamma_{3\downarrow}$ and $\gamma_{4\uparrow}$ spin components as the maximal probabilities of  admixtures of the dot states: $L_{\downarrow}=|a_{3,1}|^{2}+|a_{3,8}|^{2}$ and $L_{\uparrow}=|a_{4,2}|^{2}+|a_{4,7}|^{2}$, respectively.  The Majorana spin polarization, dependent  on this localization, assumes then the from: $P_{L}=(L_{\uparrow}-L_{\downarrow})/(L_{\uparrow}+L_{\downarrow})$.

\begin{figure} [ht]
\epsfxsize=0.3\textwidth
\epsfbox{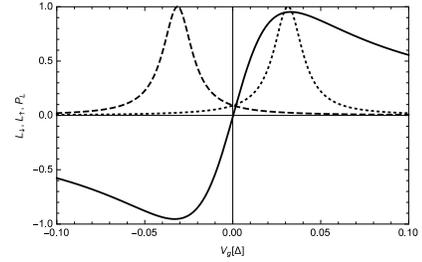}
\caption{\label{MBSloc} Majorana bound state localizations and its spin polarization vs. gate voltage: $L_{\downarrow}$-dashed, $L_{\uparrow}$-dotted and $P_{L}$-solid curve. Calculations were performed for $\Theta=0.99\pi$, $\Delta\phi=\pi$, $B=2$, $t_{L}=t_{R}=0.1$, $\epsilon_{L}=0.02$ and $\epsilon_{R}=0$.  }
\end{figure}

In Fig.~\ref{MBSloc} gate voltage dependencies of these quantities, calculated for $\Theta=0.99\pi$, are presented. The maximum of the $L_{\downarrow}$ ($L_{\uparrow}$) localization of the MBS spin component inside the dot is achieved when $\tilde{\epsilon}_{\downarrow}$  ($\tilde{\epsilon}_{\uparrow}$) dot sublevel crosses Fermi energy. Also in these regions the maxima of spin polarization appear as a result of competition between $L_{\uparrow}$ and $L_{\downarrow}$.
The polarization assumes zero value for $V_{g}=0$, when there is a symmetric arrangement of the dot spin-sublevels with respect to Fermi energy. It also approaches zero value for large $|V_{g}|$, when both the dot spin-sublevels are shifted away from Fermi energy, and the admixture of extended fermionic $f_{L}$ state dominates.

The spin polarization  of the unpaired Majorana state exhibits also $4\pi$ periodicity as a function of $\Delta\phi$ (not shown), which is a characteristic feature of Josephson junction with an MBS.

The admixture of the fermionic states to the unpaired $\gamma^{R}_{A,1}$ MBS  can be considered as Majorana "leaking"  through the dot into the left wire. It can be switched off in a given spin sector by aligning the corresponding   QD spin-sublevel with Fermi level. Similar "leaking" of the Majorana wave function from the end of topological wire into a quantum dot in side-coupled geometry has been predicted \cite{Vernek} and also observed experimentally \cite{Deng3}. In the side-coupled geometry the maximal "leakage" has been observed when the dot level was tuned to resonance with the MBS.

\subsection{Josephson current and spin polarization of ABS levels}

Diagonalization of the Hamiltonian written in the Nambu basis reveals the ABS spectrum of the junction.
The dc phase-biased Josephson current, carried by the quasiparticle with energy $E_{i}$, reads \cite{Zagoskin}:
\begin{equation}
\label{current0}
J_{i}=\frac{2e}{\hbar}\frac{\partial E_{i}}{\partial\Delta\phi}
\end{equation}
In our case $E_{i}$ correspond to the Andreev levels arising in the dot region. Andreev levels, as particle-hole states, have no well defined parity. In the further analysis we do not rely on the conservation of parity and assume a scenario of thermodynamical equilibrium; that the population inversion of the Andreev levels after their crossing is not necessarily conserved. The current in this case can be written as a sum over the available Andreev levels with appropriate weights defined by Fermi distribution function:
\begin{equation}
\label{current}
J_{\phi}=\frac{2e}{\hbar}\sum_{i}\frac{\partial E^{ABS}_{i}}{\partial\Delta\phi}f(E^{ABS}_{i}).
\end{equation}
It is also instructive to analyze the spin polarization, Eq.~(\ref{spin_pol}), which can be ascribed to the $i$-th ABS level. In this case
the generalized particle operator is constructed from the operators, Eq.~(\ref{ABSoperator}), and the matrix elements are calculated between the states of the Hilbert space generated by $|n\rangle_{L} \otimes |n\rangle_{d}\otimes |n\rangle_{R}$ product with the maximal spin up or down. We also define an effective thermally averaged polarization of the junction:
\begin{widetext}
\begin{equation}
P_{therm}=\frac{\sum_{i}f(E_{i}^{ABS})[\langle\uparrow\uparrow\uparrow|n_{i}|\uparrow\uparrow\uparrow\rangle-\langle\downarrow\downarrow\downarrow|n_{i}|\downarrow\downarrow\downarrow\rangle]}
{\sum_{i}f(E_{i}^{ABS})[\langle\uparrow\uparrow\uparrow|n_{i}|\uparrow\uparrow\uparrow\rangle+\langle\downarrow\downarrow\downarrow|n_{i}|\downarrow\downarrow\downarrow\rangle]}
=\frac{\sum_{i}f(E_{i}^{ABS})P_{i}}{\sum_{i}f(E_{i}^{ABS})},
\end{equation}
\end{widetext}
in which the polarizations of the subsequent ABS levels are weighted by their corresponding occupancies in a given temperature. This quantity is more likely to be experimentally detectable.

\subsection{Paired MBS for $\epsilon_{L}$, $\epsilon_{R}\neq 0$}
Let us examine the most general case, when all the MBS are paired by the finite hybridization in each wire. It follows from  diagonalization of the Hamitonian matrix  that, independently of the value of $\Theta$, there are two Majorana bound states localized at Fermi level. Their spin components are described in terms of extended Dirac fermions residing in the left and the right wire: $\gamma_{1\sigma}=(1/\sqrt{2})(f^{\dagger}_{R\sigma}+f_{R\sigma})=(1/\sqrt{2})\gamma^{R}_{A,1\sigma}$ and $\gamma_{2\sigma}=(i/\sqrt{2})(f^{\dagger}_{L\sigma}-f_{L\sigma})=(1/\sqrt{2})\gamma^{L}_{B,N\sigma}$. They have overall zero spin polarization. This result can be compared to the case of $\Theta=0$, when spin-down sector is only active and the MBS have $P_{i}=-1$.

\begin{figure*} [ht]
\epsfxsize=0.65\textwidth
\epsfbox{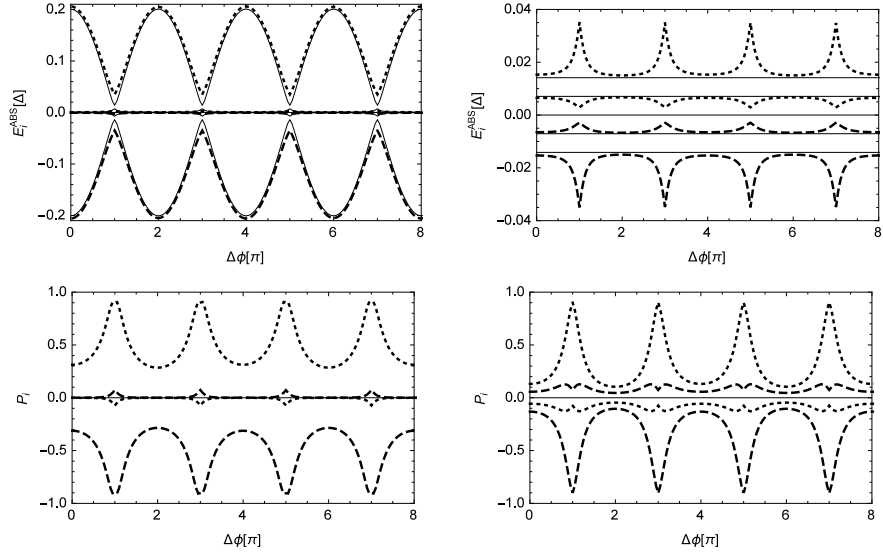}
\caption{\label{ABSandPV0Th099} Upper panels: energy structure of the junction. Thin solid curves are the ABS levels calculated for $\Theta=\pi$, and the dashed and dotted curves are the pairs $E^{ABS}_{i\mp}$ for $\Theta=0.99\pi$. In the upper left panel: the saw-tooth-like levels with strong $\Delta\phi$ dependence are $E^{ABS}_{5\mp}$ and diamond-like levels close to $\epsilon_{F}$ are $E^{ABS}_{6\mp}$. In the upper right panel the $E^{ABS}_{8\mp}$ pair closer to $\epsilon_{F}$ and $E^{ABS}_{7\mp}$ pair further away do not depend on $\Delta\phi$ for $\Theta=\pi$. The corresponding pairs of levels for $\Theta=0.99\pi$ are shown by dashed and dotted curves. Lower panels: spin polarization of the ABS levels shown in the upper panels with the same notation as the corresponding ABS levels. For $\Theta=\pi$ the ABS levels are unpolarized.  Calculations were performed for $B=2$, $t_{L}=t_{R}=0.1$, $\epsilon_{L}=0.02$ and $\epsilon_{R}=0.01$, $V_{g}=0$.}
\end{figure*}

Apart from the MBS states, the energy spectrum of the junction is consists of four pairs of $E^{ABS}_{\mp i}$ Andreev levels.
In Fig.~\ref{ABSandPV0Th099} the phase evolution of these levels (upper panels) and their spin polarizations (lower panels) are shown, calculated for the finite Zeeman splitting in the dot ($\Theta=0.99\pi$) and particle-hole symmetry ($V_{g}=0$). The levels display $2\pi$ periodicity in $\Delta\phi$, which does not reveal the MBS existence. The $4\pi$ periodicity, characteristic for the MBS,  would emerge if the parity were conserved in the junction. This condition is difficult to achieve experimentally due to various quasi-particle poisoning processes \cite{Fu&Kane2,Lutchyn,Badiane,Jiang,Pikulin}. Nevertheless, as shown \cite{PStefanskiJPCM16}, the $4\pi$ periodicity can be revealed upon breaking an overall p.-h. symmetry of the junction, by setting the gate voltage  $V_{g}\neq 0$ (not shown here). Similar structure of the Andreev levels and their phase dependence have been obtained within more realistic microscopic models of the Josephson junction \cite{sanjose,cayao}, which supports the validity of the results obtained within the present low energy effective model.

Two kinds of the ABS levels can be distinguished with respect to their dependence on superconducting phase difference: i) saw-tooth-like levels, exhibiting strong $\Delta\phi$-dependence and ii) diamond-like levels, exhibiting weak  $\Delta\phi$-dependence and lying close to Fermi energy. The former levels are dominated by the admixture of QD $d$-states, whereas the latter levels are dominated by the extended fermionic  $f_{L}$ and $f_{R}$ state contributions, and are situated at the energies $\sim\mp\epsilon_{\alpha}$, $\alpha=L,R$. These different dependencies of the ABS levels on $\Delta\phi$ can be correlated with the spatial localization of their dominant contributions. The sensitive levels have the dominant QD $d$-state contribution, which is localized in the middle of the junction, where the phase bias is the strongest. The less sensitive ABS have dominant contributions from extended fermions $f_{\alpha}$ localized in the wires, where the phase bias is diminished.

As shown in the upper right panel of Fig.~\ref{ABSandPV0Th099}, for $\Theta=\pi$ some of the ABS levels become $\Delta\phi$-independent. This feature can be understood in terms of superposition of the particle and hole tunneling through the dot spin sublevels: $\tilde{\epsilon}_{\sigma}$ and $-\tilde{\epsilon}_{-\sigma}$, respectively, see the Hamiltonian matrix, Eq.~(\ref{H_matrix}). A part of the ABS levels depend on the sum of particle and hole tunneling $\tilde{\epsilon}_{\uparrow}-\tilde{\epsilon}_{\downarrow}$ and for the spin degeneracy (for $\Theta=\pi$) the renormalization terms, containing  phase dependence of these processes, cancel out leaving this class of ABS levels $\Delta\phi$ independent.

The extremes of the levels appearing for the odd values of $\Delta\phi$ are correlated with the position of the renormalized dot level: for $\Theta=\pi$ they appear for $\tilde{\epsilon}_{\downarrow}=\tilde{\epsilon}_{\uparrow}=\epsilon_{F}$ and for finite magnetic field for $(\tilde{\epsilon}_{\downarrow}+\tilde{\epsilon}_{\uparrow})/2=\epsilon_{F}$.

In the lower panels spin polarizations ascribed to the pairs of ABS levels  are shown. They are in relation $P_{i+}=-P_{i-}$, which demonstrates that each Andreev reflection process is composed of a tunneling electron with spin $\sigma$ and an accompanying reflected hole with spin $-\sigma$. With the increase of the magnetic field (not shown) the saw-tooth-like levels $E^{ABS}_{5\mp}$, shown in upper left panel, gradually shift apart from Fermi energy, whereas the diamond-like levels  $E^{ABS}_{6\mp}$, lying close to Fermi energy, approach $\epsilon_{F}$ even further. Similar behavior of the ABS levels depicted in the upper right panel can be observed: the pair of levels, $E^{ABS}_{7\mp}$, lying farther from $\epsilon_{F}$, spread further apart on the energy scale, and the levels  $E^{ABS}_{8\mp}$, lying closer to $\epsilon_{F}$, approach Fermi energy with the increase of the effective magnetic field in the dot. The polarizations of the ABS pairs of levels spreading away from Fermi energy when magnetic field increases, gradually approach the value $P_{i\mp}=\mp 1$ with diminished oscillations, whereas the polarizations of the ABS pairs approaching $\epsilon_{F}$ gradually decrease to zero.
For $\Theta=\pi$, spin polarization of both the particle-like and the hole-like ABS levels is zero, $P_{i\mp}=0$. This is the result of equal spin-up and spin-down operator contributions in the quasiparticle operator, Eq.~(\ref{ABSoperator}), realized for zero effective magnetic field in the junction.

\begin{figure} [ht]
\epsfxsize=0.5\textwidth
\epsfbox{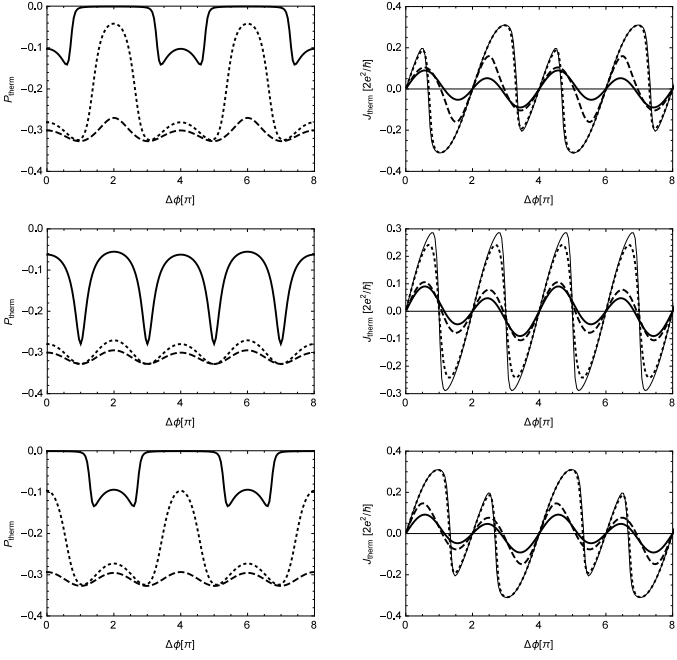}
\caption{\label{ThermalPandJ} Thermally averaged spin polarizations of the junction (left column) and the Josephson current (right column) calculated for $V_{g}=-0.1$ - upper panels, $V_{g}=0$ - middle panels and $V_{g}=0.1$ - lower panels. Solid curves correspond to $\Theta=0.99\pi$, dotted curves correspond to $\Theta=0.95\pi$ and dashed curves correspond to $\Theta=0.93\pi$. The thin solid curve in the right column represents thermally averaged Josephson current calculated for $\Theta=\pi$. Calculations were performed for $B=2$, $t_{L}=t_{R}=0.1$, $\epsilon_{L}=0.02$ and $\epsilon_{R}=0.01$ and T=0.01.}
\end{figure}

In Fig.~\ref{ThermalPandJ}  thermally averaged spin polarizations of the junction and the corresponding effective Josephson currents are shown, calculated for various gate voltages and effective magnetic fields in the dot. Since these quantities are weighted by Fermi distribution function, they are dominated mainly by the lowest-lying ABS levels, in particular $E^{ABS}_{i-}$ level (see Fig.~\ref{ABSandPV0Th099}), which has negative spin polarization. For $V_{g}=0$, when the p.-h. symmetry is conserved, a clear $2\pi$ periodicity in $\Delta\phi$ of the Josephson current  can be observed. However, with increasing magnetic field inside the dot, the $4\pi$ periodicity emerges even for $V_{g}=0$ (see the solid curve for $\Theta=0.93\pi$). This is caused by an increase of the Zeeman splitting of the dot spin-sublevels and the thermally  promoted dominance of occupancy of the lower lying $\epsilon_{\downarrow}$ level at the expense of the  higher lying $\epsilon_{\uparrow}$ level occupancy. Also the ABS levels carrying current through the junction become dependent on non-evenly occupied particle and hole channels (with opposite spins), which introduces spin polarization.

In general, as seen in the right column, the increasing magnetic field inside the dot diminishes the Josephson current oscillations. For $\Theta=\pi$ the term describing spin mixing in the tunneling Hamiltonian, Eq.~(\ref{H_tun_simp_fer}), acquires the largest value, and also  the amplitude of the tunneling supercurrent is maximal.  For such an arrangement  there is a perfect equilibrium between spin-up and spin-down electrons inside the dot; the tunneling electron with spin $\sigma$ can easily pair with the reflected hole with opposite spin, and form the ABS level. For the finite magnetic field, the balance between the spin sectors is broken; thus it is more difficult to find the $-\sigma$ hole and to form an ABS with $\sigma$ electron. One notes that this process is sensitive to  effective magnetic field in the dot.

For $V_{g}\neq 0$, when the p.-h. symmetry is broken, clear $4\pi$-periodic oscillations can be observed in both the polarizations and the currents.

\subsection{Spin polarization of the junction for $\epsilon_{L}\neq 0$ and $\epsilon_{R}=0$}
\begin{figure} [ht]
\epsfxsize=0.5\textwidth
\epsfbox{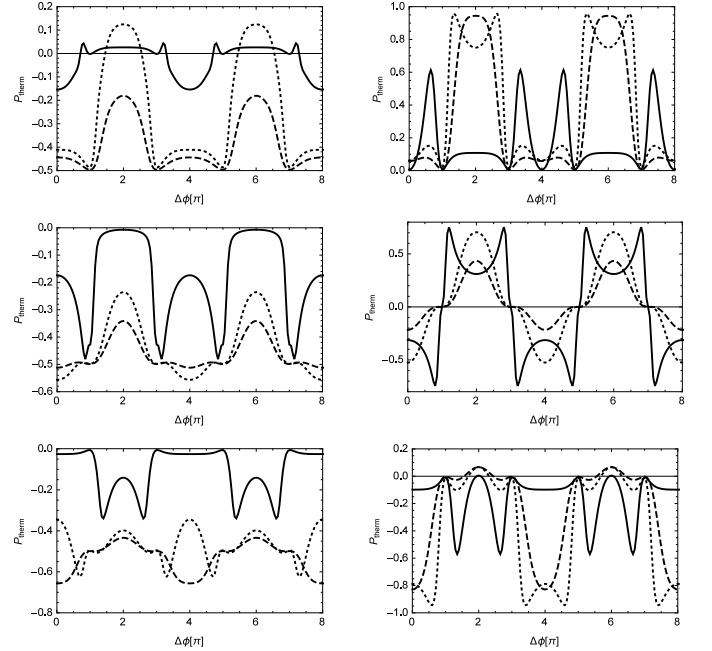}
\caption{\label{thermalPandPofMBSlike} Thermally averaged spin polarizations of the junction (left column) and the contribution to spin polarizations coming from unpaired MBS  (right column) calculated for $V_{g}=-0.1$- upper Panels, $V_{g}=0$- middle Panels and $V_{g}=0.1$- lower Panels. Solid curves are for $\Theta=0.99\pi$, dotted curves are for $\Theta=0.95\pi$ and dashed curves are for $\Theta=0.93\pi$. Calculations were made for $B=2$, $t_{L}=t_{R}=0.1$, $\epsilon_{L}=0.02$ and $\epsilon_{R}=0$ and T=0.01.}
\end{figure}

For the finite hybridization of the MBS in the left wire, there is a paired Majorana bound state with zero spin polarization and an unpaired MBS with admixtures of extended fermion $f_{L\sigma}$, Eqs.(\ref{Maj1})-(\ref{Maj4}). The other ABS levels left behave similarly as $E^{ABS}_{5\mp}$ and $E^{ABS}_{7\mp}$, shown in Fig.~\ref{ABSandPV0Th099}. As a consequence, the Josephson current in thermal equilibrium has similar phase dependencies as shown in Fig.~\ref{ThermalPandJ}. However, the thermally averaged spin polarization of the junction exhibits a different behavior due to the influence of  the unpaired MBS with the spin contributions of Eqs.~(\ref{Maj3}) and (\ref{Maj4}). This feature is demonstrated in Fig.~\ref{thermalPandPofMBSlike}, where in the left column  thermally averaged spin polarizations of the junction vs. $\Delta\phi$ are presented and in the right column the contribution originating from the unpaired MBS.
Dependent on the position of the QD level, this contribution can be positive (for $V_{g}<0$), negative (for $V_{g}>0$) or in the range $P_{therm}\simeq\mp 1$ for particle-hole symmetry ( $V_{g}=0$). As compared to the fully paired Majorana case, shown in Fig.~\ref{ThermalPandJ}, where $P_{therm}$  assumed negative values, in the present case there can be a sign switching of the overall polarization due to the positive contribution from unpaired MBS. This switching is $4\pi$-periodic in $\Delta\phi$. Moreover, due to the presence of unpaired MBS, the thermally averaged spin polarization remains $4\pi$-periodic even when p.-h. symmetry is not broken (see the curves for  $V_{g}=0$). This is in contrast to the fully paired MBS case, shown in the left middle panel of Fig.~\ref{ThermalPandJ}, where $2\pi$ periodicity can be observed. This $4\pi$ periodic polarization switching can be utilized as a tool for searching of the Majorana states in Josephson junctions without relying on the parity conservation.

\section{Conclusions}

We considered a low energy, effective model of a Josephson junction of two topological wires and mediated by an interacting quantum dot. The Majorana states adjacent to the dot hybridize across the junction and form a QD bound state. The dot is exposed to an effective magnetic field created by the fields driving the wires into topological state. As a result of finite Zeeman splitting of the QD bound state, the spin components of the Majorana states can be analyzed separately in the presence of electron interactions in the dot. We showed that electron correlations diminish the ZBC Majorana peak from a half of conductance quantum for a non-interacting dot to $\sim(2/5)e^{2}/h$. For shorter wires, when the overlap of the end-state Majorana wavefunctions cannot be neglected, electron interactions also renormalize this hybridization, which becomes dependent on the dot parity and generally is diminished.

Due to the specific geometry of the device, the unpaired Majorana "leaking" through the dot  into the opposite wire with paired MBS can be observed, which is spin-dependent and can be controlled by tuning the position of the QD level with respect to Fermi energy.

Josephson current in thermal equilibrium exhibits variable spin polarization dependent on the resultant Zeeman field in the dot. Moreover, it possesses $4\pi$ periodicity vs. phase bias characteristic for Majorana-assisted tunneling, which does not require  experimentally challenging conservation of parity of the junction. Also in the presence of the unpaired MBS in the junction, the thermally averaged spin polarization of the current alters its sign with the period $4\pi$ vs. phase bias. The control of the Josephson current spin polarization can potentially be applied in spintronic applications.



\end{document}